%% file: main.tex
\documentclass[journal,comsoc, 12pt, onecolumn, draftclsnofoot]{IEEEtran}
\usepackage{cite}
\usepackage{amsmath,amssymb,amsfonts,helvet}
\usepackage[cmintegrals]{newtxmath}
\usepackage{bbm}
\usepackage{algorithmic}
\usepackage{graphicx}
\usepackage{textcomp}
\usepackage{soul}
\def\BibTeX{{\rm B\kern-.05em{\sc i\kern-.025em b}\kern-.08em
    T\kern-.1667em\lower.7ex\hbox{E}\kern-.125emX}}
    
\usepackage[
  pass,
]{geometry}
\usepackage{graphicx,epsfig}
\usepackage{subfig}
\usepackage{caption,setspace}
\usepackage{cite}
\usepackage{url}

\usepackage{footmisc}
\usepackage[alwaysadjust]{paralist}
\usepackage{hyperref}
\usepackage{multirow}
\usepackage{color, colortbl}
\usepackage{soul}
\usepackage{listings}
\usepackage{color}
\usepackage{makecell}

\usepackage{acro}
\input{acronyms.tex} 

\def\argmax{\mathop{\rm arg\,max}}

\begin{document}

\title{Learning Site-Specific Probing Beams for Fast mmWave Beam Alignment}

\author{Yuqiang~Heng,
        Jianhua~Mo
        and~Jeffrey~G.~Andrews,~\IEEEmembership{Fellow,~IEEE}
\thanks{Yuqiang Heng and Jeffrey G. Andrews are with the Wireless Networking and Communications Group (WNCG), The University of Texas at Austin, Austin, TX 78701 USA. Email: (yuqiang.heng@utexas.edu, jandrews@ece.utexas.edu).}
\thanks{Jianhua Mo is with Samsung Research America, Plano, TX 75023, USA. Email: jianhua.m@samsung.com}
\thanks{A conference version of this work is submitted to IEEE GLOBECOM 2021 \cite{ConferenceVersion}.}
}



\maketitle

\begin{abstract}
Beam alignment -- the process of finding an optimal directional beam pair -- is a challenging procedure crucial to \ac{mmWave} communication systems. We propose a novel beam alignment method that learns a site-specific probing codebook and uses the probing codebook measurements to predict the optimal narrow beam. An end-to-end \ac{NN} architecture is designed to jointly learn the probing codebook and the beam predictor. The learned codebook consists of site-specific probing beams that can capture particular characteristics of the propagation environment. The proposed method relies on beam sweeping of the learned probing codebook, does not require additional context information, and is compatible with the beam sweeping-based beam alignment framework in 5G.  Using realistic ray-tracing datasets, we demonstrate that the proposed method can achieve high beam alignment accuracy and \ac{SNR} while significantly -- by roughly a factor of 3 in our setting -- reducing the beam sweeping complexity and latency.    
\end{abstract}

\begin{IEEEkeywords}
5G mobile communication, Beam steering, Beam management, Beam codebook, Machine learning, Millimeter wave communication, Supervised learning.
\end{IEEEkeywords}
\IEEEpeerreviewmaketitle

\section{Introduction}\label{section:introduction}
Cellular systems will increasingly tap into the \acf{mmWave} spectrum to provide higher data rates and to support a wide range of emerging use cases. For example, the current release of 5G adopts several mmWave bands between 24.25 GHz and 52.6 GHz, while future releases are expected to further expand the spectrum to 71 GHz and even the so-called ``Terahertz'' bands extending up to 300 GHz \cite{3gpp.NR.above52GHz}. While these high carrier frequencies allow much larger bandwidths, they also impose harsher propagation conditions and utilize large arrays of very small antenna elements, and thus rely on highly directional \ac{BF} to maintain viable received signal strength. Meanwhile, these directional links are highly sensitive to blockage and reflections, so beam alignment -- finding and maintaining near-optimal analog \ac{BF} weights, including for \ac{NLOS} paths -- is essential.  MmWave devices typically adopt codebooks of indexed analog beams to allow good beams to be identified by the receiver and fedback to the transmitter. These codebooks will contain much more numerous and much narrower beams as higher carrier frequencies are adopted, making the latency and beam sweeping overhead of traditional beam searches prohibitive. As a result, beam alignment will become an increasingly critical bottleneck in the future. 

\subsection{Background and Related Work} \label{section:related_work}
\begin{figure*}[hbt]
   \centering
   \includegraphics[width=0.95\textwidth]{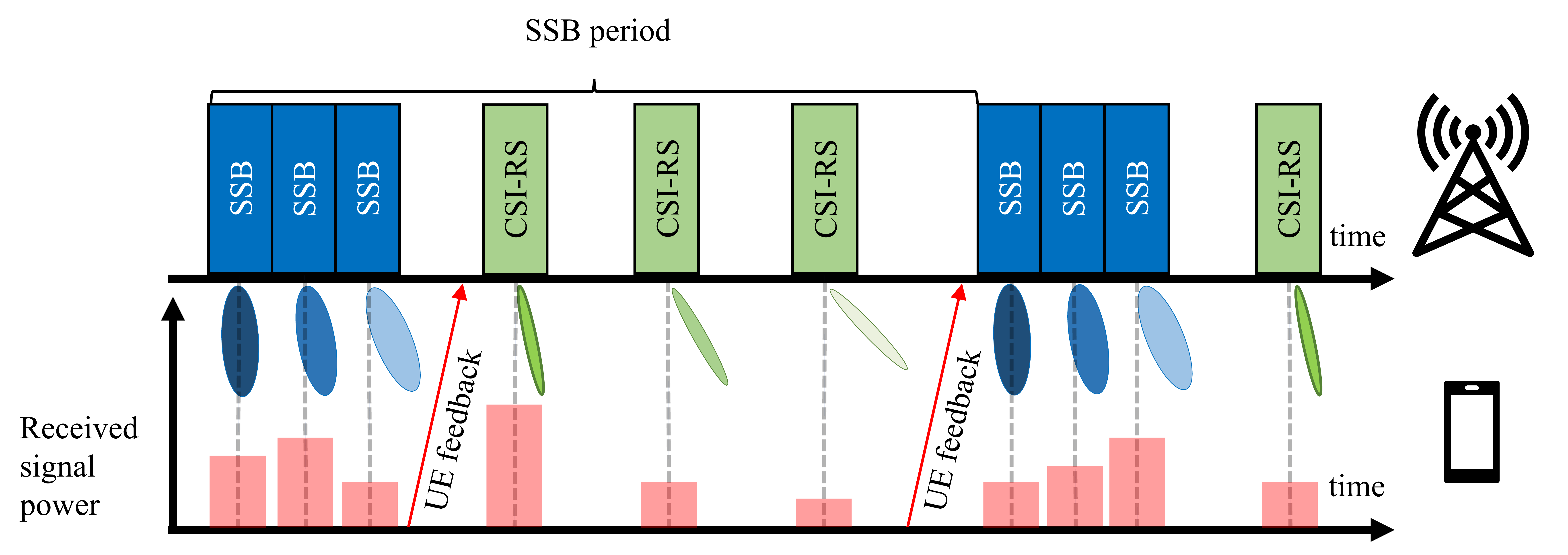}
   \caption{Beam sweeping-based beam alignment framework in 5G. The \ac{BS} transmits beamformed \acp{SSB} and \acp{CSI-RS}. The \ac{UE} measures and reports the quality of the \acp{RS}. The \ac{BS} selects a beam for future data or control transmission.}
   \label{fig:5G_BM_RS}
\end{figure*}

The current release of 5G adopts a beam alignment framework based on beam sweeping, measurements and reporting \cite{Giordani19_BM_Tutorial},\cite{Li20_BM_NR},\cite{heng2020BM_magazine}. In the \ac{DL}, the \ac{BS} transmits \acp{RS} such as \acp{SSB} and \acp{CSI-RS} using different beams to sweep the angular space, as illustrated in Fig. \ref{fig:5G_BM_RS}. The \ac{UE} uses a quasi-omnidirectional beam or sweeps its beam codebook using different receiving beams, measures the receive signal power, then reports the RS measurements to the BS. With exhaustive beam sweeping, the \ac{BS} and the \ac{UE} need to search all combinations of beam pairs, resulting in significant beam sweeping overhead and latency. 
The \acp{SSB} are transmitted periodically and are ``always-on''. They are also used in cell discovery and \ac{IA} for new \acp{UE}. In order for an unconnected UE to achieve synchronization before accessing the network, it needs to measure the \acp{SSB} transmitted by the \ac{BS}, find one associated with a good beam and derive the necessary information from that \ac{SSB}. Since beam sweeping is essential for both beam alignment and cell search, a beam sweeping-based framework is likely going to stay in future releases of 5G.

Hierarchical beam searches have been proposed to reduce the beam sweeping complexity \cite{Hierarchy01},\cite{Compare01}. The \ac{BS} and the \ac{UE}, equipped with multiple-tier codebooks, sweep wider beams first and iteratively thin the search space for the best narrow beam. Since mmWave systems often employ analog or hybrid \ac{BF}, the hardware constraints need to be considered when designing the wide beams in these hierarchical codebooks. Different hierarchical codebook design techniques have been proposed in recent works such as \cite{xiao2016hierarchical} and \cite{qi2020hierarchical}. While the hierarchical search reduces the number of beams swept  compared to an exhaustive one, the search procedure needs to be repeated for each UE, marginalizing the gain for multiple UEs. They are also more susceptible to search errors caused by noise in received signal and imperfect wide-beam patterns. 
The hierarchical search method proposed in \cite{qi2020hierarchical} uses wide beams with multiple mainlobes to reduce the beam sweeping overhead for multiple UEs. However, the intermediate-layer beams need to be dynamically generated based on measurements of upper-layer beams, which lacks standardization support from 5G and also significantly increases the size of the effective hierarchical codebook.

In addition to the beam sweeping-based approaches, beam alignment methods that utilize context information has been explored. In \cite{wang2018mmwave},\cite{heng2021MLbeamalignment} and \cite{va2017inverse}, the location information of UEs are used to reduce the beam search space. Beam alignment methods that utilizes sub-6 GHz measurements are proposed in \cite{ali2018oob}, \cite{alrabeiah2020sub6ghz} and \cite{nitsche2015oob}. In \cite{alkhateeb2018deep}, omni-directionally received sounding signals are used to predict the optimal beam. A beam alignment method assisted by radar measurements is proposed in \cite{nuria2016radar}. However, such context information can be hard to obtain since mmWave devices need to be equipped with the required additional sensors. The feedback of such context information also incurs additional overhead and sometimes requires a more robust sub-6 GHz link between the \ac{BS} and the \ac{UE}.

\Ac{ML} solutions have been explored for the beam alignment problem. The pattern extraction and function approximation powers of ML models make them particularly suitable for processing a wide range of context information, such as location \cite{heng2021MLbeamalignment}, sub-6 GHz channels \cite{alrabeiah2020sub6ghz} and omni-directional sounding signals \cite{alkhateeb2018deep}. A joint BF, power control and interference coordination method using reinforcement learning (RL) is proposed in \cite{mismar20RLbeamforming}. A beam alignment method that uses \ac{CS} to leverage channel sparsity is proposed in \cite{myers2019cs_beamalignment}. In \cite{myers2020deeplearning_cs_beamalignment}, a deep learning architecture is used to learn CS matrices and predict the best beams.   

Compared to the beam sweeping-based approaches, beam alignment solutions that rely on context information often require an additional cell search procedure to discover unconnected new \acp{UE} regardless of whether traditional \ac{ML} or deep learning techniques are used. The feedback of additional context information requires the \ac{UE} to be connected to the network through mmWave links or sub-6 GHz side links, which can be problematic during \ac{IA}. Furthermore, solutions that do not adopt beam sweeping are not compatible with the beam alignment framework in 5G. Significant modifications to the 5G standard is required to accommodate these approaches.

The \acf{NN} architecture proposed in this work consists of a complex layer which represents the analog beam codebook and an \ac{MLP} which acts as the beam selector. The complex-\ac{NN} layer used in this work was first proposed in \cite{trabelsi2017deepcomplexNN} for computer vision and audio-related tasks. A similar complex fully-connected layer is used in \cite{alrabeiah2020NNcodebook_arxiv}, where the authors optimize beam patterns for particular environments and hardware imperfections. This work focuses on finding an optimal beam from a large predefined narrow-beam codebook and differs from \cite{alrabeiah2020NNcodebook_arxiv} which focuses on direct codebook learning. 

In \cite{ma2020MLbeamalignment}, a beam alignment method that trains a \ac{NN} to predict optimal beams using \ac{UL} measurements from a sparse probing codebook is proposed. However, the probing codebooks used in \cite{ma2020MLbeamalignment} are predetermined undersampled \ac{DFT} codebooks with evenly spaced narrow beams, whereas the probing codebook in our proposed method are site-specific and learned using a complex-valued \ac{NN} module. The \ac{NN} in \cite{ma2020MLbeamalignment} requires knowledge of the complex received signals, whereas our proposed method only need the received power. We demonstrate in Section \ref{section:eval_explanation} that our learned probing codebooks are much more effective at capturing characteristics of the environment and providing useful information to the beam predictor.

\subsection{Contributions}
In this work, we propose a beam alignment method that uses the beam sweeping measurements of a probing codebook to predict the optimal narrow beam. The proposed method is based on beam sweeping and does not require any additional context information, which is compatible with the beam alignment framework in 5G. By jointly training the probing codebook and the beam predictor using a \ac{NN} in an end-to-end fashion, the probing codebook is able to learn particular characteristics of the propagation environment and optimize its beam patterns accordingly.  Some key features of the proposed method are summarized as follows. 

\textbf{Trainable site-specific probing codebook:} A complex-\ac{NN} module is used to parameterize the probing codebook during training so that the \ac{BF} weights can be extracted and implemented using actual \ac{RF} chains during deployment. The probing codebook is able to learn particular characteristics of the propagation environment and optimize its beams to capture the channel information effectively. The proposed architecture can be adopted by \acp{BS} in various deployment scenarios with arbitrary array geometry. 

\textbf{Compatibility with 5G framework:} The proposed method can be directly adopted without modifications to the 5G standard. It does not require the collection and feedback of hard-to-obtain context information such as UE location or out-of-band information, which needs additional standardization support. Instead, the proposed method uses beam sweeping measurements of a probing codebook, which is exactly compatible with the beam sweeping-based framework currently adopted in 5G. The probing beams can be transmitted using \acp{SSB}, which can also be used for cell discovery and \ac{IA}.

\textbf{High beam alignment accuracy and SNR:} We demonstrate using multiple realistic ray-tracing datasets that the proposed method can achieve high beam alignment accuracy and \acf{SNR}, beating the hierarchical beam search baselines. 
For instance, the proposed method can achieve a beam alignment accuracy of over 90\% and can outperform even the exhaustive search in terms of the average \ac{SNR}.

\textbf{Reduced beam sweeping overhead:} The proposed method has lower beam sweeping overhead compared to exhaustive and hierarchical beam searches, especially when considering beam alignment for multiple \acp{UE}. For instance, when considering simultaneous beam alignment for 10 \acp{UE}, the proposed method is about 3$\times$ faster compared to exhaustive and hierarchical beam searches.

\textbf{Applicable to a wide range of propagation scenarios:} Multiple accurate ray-tracing datasets modelling a wide range of propagation environments are used to evaluate the performance of the proposed method. The proposed beam alignment approach consistently achieves high accuracy and \ac{SNR} in indoor and outdoor environments, for \ac{LOS} and \ac{NLOS} \acp{UE}, and with 28 GHz and 60 GHz carrier frequencies.

The rest of this article is organized as follows. The system model is described in Section \ref{section:system_model}. The proposed beam alignment approach, the appropriate metrics and the baselines of comparison are explained in Section \ref{section:proposed_method}. The datasets used are described in Section \ref{section:dataset}. The simulation results are presented in Section \ref{section:evaluation}. Finally, the conclusion and final remarks are provided in Section \ref{section:conclusion}.

\section{System Model}\label{section:system_model}
A \ac{DL} \ac{MISO} system is considered, where each \ac{BS} has an antenna array of $N_t$ elements and each \ac{UE} has a single antenna. While \acp{UE} typically have antenna arrays also, we consider a \ac{MISO} scenario where beam alignment is only performed on the \ac{BS} side for simplicity. 
The \ac{MISO} model is also applicable to the \ac{mMTC} use case of 5G, where the each sensor would likely use a single antenna and an isotropic beam pattern.
The extension to receive beam alignment on the \ac{UE} side is left to future work.  
A ray-based narrowband block-fading \ac{mmWave} channel model with $N_p$ paths is considered \cite{heath2016overview}:
\begin{align}\label{eqn:channel_model}
    \mathbf{h} = \sum_{l=1}^{N_p} \alpha_l\mathbf{a}(\phi_l^D, \theta_l^D).
\end{align}
For each path $l$, its complex gain is $\alpha_l$, the azimuth and elevation angles of departure are $\phi_l^D$ and $\theta_l^D$, and the array steering vector at these angles is denoted by $\mathbf{a}(\phi_l^D, \theta_l^D)$. For a \ac{ULA} with $N_{t}$ antenna elements on the $y$-axis, its beam steering is limited to the azimuth domain and its steering vector can be written as 
\begin{equation}
\mathbf{a_{ULA}}(\phi_{l}) = \frac{1}{\sqrt{N_t}} 
\begin{bmatrix}
1 & e^{j\frac{2 \pi}{\lambda}d\sin \phi_{l}} & \cdots & e^{j(N_{t}-1)\frac{2 \pi}{\lambda}d\sin \phi_{l}}
\end{bmatrix}^{T},
\end{equation}
where $\lambda$ is the carrier wavelength and \textit{d} is the antenna spacing \cite{Balanis01}. 
While a \ac{ULA} is considered instead of a planar array for simplicity, the proposed beam alignment approach is array-geometry agnostic and can be applied to arrays of arbitrary geometry, as we will explain in Section \ref{section:complex_NN_module}.

Due to the cost and complexity of fully digital \ac{BF} at \ac{mmWave} frequencies, each \ac{BS} is assumed to perform analog-only or hybrid \ac{BF}. For the purpose of beam alignment, only the \ac{RF} domain processing is considered. For a \ac{BS}-\ac{UE} pair, the \ac{BS} is assumed to employ a single \ac{RF} chain to which all antenna elements are connected. Analog \ac{BF} is assumed to be implemented using phase shifters connected to each antenna element. The \ac{BF} vector can be written as 
\begin{equation}
\mathbf{v} = \frac{1}{\sqrt{N_t}}
\begin{bmatrix}
e^{j\theta_1} & e^{j\theta_2} & \cdots & e^{j\theta_{N_{t}}} 
\end{bmatrix}^T,
\end{equation}
where $\mathbf{v}$ satisfies the power constraint and each element of $\mathbf{v}$ satisfies the constant modulus constraint.

In the \ac{DL}, the \ac{BS} transmits a symbol $s \in \mathbb{C}$ satisfying average power constraint $\mathbb{E}[|s|^2] = 1$ to the UE using a \ac{BF} vector $\mathbf{v}$. The received signal at the \ac{UE} can be written as
\begin{equation}\label{eq:bf_signal}
y = \sqrt{P_T}\mathbf{h}^H\mathbf{v}s + n,   
\end{equation}
where $P_T$ is the transmit power, $\mathbf{h} \in \mathbbm{C}^{N_{t} \times 1}$ is the channel vector and $n$ is the complex additive noise with noise power $\sigma_n^2$.

The \ac{SNR} for a \ac{UE} with channel $\mathbf{h}$ and using a \ac{BF} vector $\mathbf{v}$ can be written as
\begin{equation}\label{eq:snr}
\text{SNR} = \frac{P_T|\mathbf{h}^H\mathbf{v}|^2}{\sigma_n^2}.
\end{equation}

The \ac{BS} has a codebook $\mathbf{V} \in \mathbbm{C}^{N_t \times N_{\mathbf{V}}}$ of predefined narrow analog beams that are used for the data or the control channel, where each column of $\mathbf{V}$ represents the \ac{BF} weights of a beam. The size of the narrow-beam codebook $N_{\mathbf{V}}$ is assumed to be large since $\mathbf{V}$ needs to cover the entire angular space.
For a \ac{BS} and a \ac{UE}, the optimal narrow beam index is the one that achieves the maximum \ac{SNR}:
\begin{equation}\label{eq:opti_beam}
i_\mathbf{v}^* = \argmax_{i \in \left\{1, 2, \cdots, N_{\mathbf{V}} \right\}}\left(\frac{|\mathbf{h}^H\mathbf{v}_i|^2 P_T}{\sigma_n^2}\right) = \argmax_{i \in \left\{1, 2, \cdots, N_{\mathbf{V}} \right\}}(|\mathbf{h}^H\mathbf{v}_i|^2),
\end{equation}
where $\mathbf{v}_{i}$ is the $i$th beam, i.e., the $i$th column of $\mathbf{V}$.

\section{The Proposed Method, Metrics and Baselines}\label{section:proposed_method}
We propose a beam alignment method that is compatible with the beam sweeping-based framework in 5G so that it can serve to achieve both beam alignment for connected \acp{UE} and \ac{IA} for unconnected \acp{UE} using only beam sweeping measurements. With the proposed method, the \ac{BS} first sweeps a small probing codebook to gather information about the channel then selects candidate narrow beams based on the probing-codebook measurements. In addition to the size-$N_{\mathbf{V}}$ narrow-beam codebook $\mathbf{V} \in \mathbbm{C}^{N_t \times N_{\mathbf{V}}}$ that is used for the data or the control channel, the \ac{BS} also has a probing codebook $\mathbf{W} \in \mathbbm{C}^{N_t \times N_{\mathbf{W}}}$ with $N_{\mathbf{W}}$ beams. The size of the probing codebook $N_{\mathbf{W}}$ is much smaller than $N_{\mathbf{V}}$. The \ac{BS} first sweeps its probing codebook $\mathbf{W}$. All \acp{UE} connected to the \ac{BS} measure and report the received power of the probing signals. The beam sweeping, measurement and reporting is assumed to be completed within the coherence time during which the channel remains the same. The reported beam sweeping results for each UE $u$ can be written as
\begin{align} \label{eq:bf_signal_power}
    \mathbf{x} &= 
\begin{bmatrix}
|y_1|^2 & \cdots & |y_{N_{\mathbf{W}}}|^2
\end{bmatrix}^T,
\end{align}
where $y_i = \sqrt{P_T}\mathbf{h}^H\mathbf{w}_{i}s + n_i$ is the received signal using the $i$th probing beam (the $i$th column of $\mathbf{W}$).
Given the reported power of received probing signals $\mathbf{x}$ of a UE, the BS then predicts the narrow beam index $i_{\mathbf{v}} \in \left\{1, 2, \cdots, N_{\mathbf{V}} \right\}$ using a function $f:\mathbf{x} \rightarrow i_\mathbf{v}$. Overall, this problem can be formulated as 
\begin{equation}\label{eq:optimization_formulation}
\begin{array}{rrclcl}
    \displaystyle \max_{\mathbf{W},f} & \multicolumn{3}{l}{{\mathop{\mathbb{E}}_{\mathbf{h} \in \mathcal{H}}}[|\mathbf{h}^H\mathbf{v}_{i_\mathbf{v}}|^2]} \\
    \textrm{s.t.} & i_\mathbf{v} & = & f(\mathbf{x}) \\
    &|\mathbf{[W]}_{i,j}| & = & \frac{1}{\sqrt{N_t}}, \forall i=1,\cdots,N_t, \forall j=1,\cdots,N_{\mathbf{W}}. \\
\end{array}
\end{equation}

The optimization problem in (\ref{eq:optimization_formulation}) is non-convex and difficult to solve due to the constant-modulus constraint of the probing BF weights and the unknown function $f$.
The proposed method is analogous to a hierarchical beam search with 2 tiers. The probing codebook is similar to the wide-beam codebook in a hierarchical search in that they both provide rough information about the channel. Unlike the hierarchical search, the probing codebook consists of beam patterns adapted to the environment, which are not limited to wide beams. In a hierarchical method, the narrow-beam selection function $f$ picks the best child beam of the best wide beam, which incurs another round of beam measurement and report. In the proposed method, the narrow-beam selection function $f$ predicts good narrow beams by intelligently utilizing measurements of all probing beams instead of using a simple heuristic, e.g., picking the narrow beams pointing to the directions of the probing beam with the best measurement. 

\subsection{The proposed NN architecture}\label{section:NN_architecture}
\begin{figure}[!tbp]
  \centering
  \includegraphics[width=0.7\textwidth]{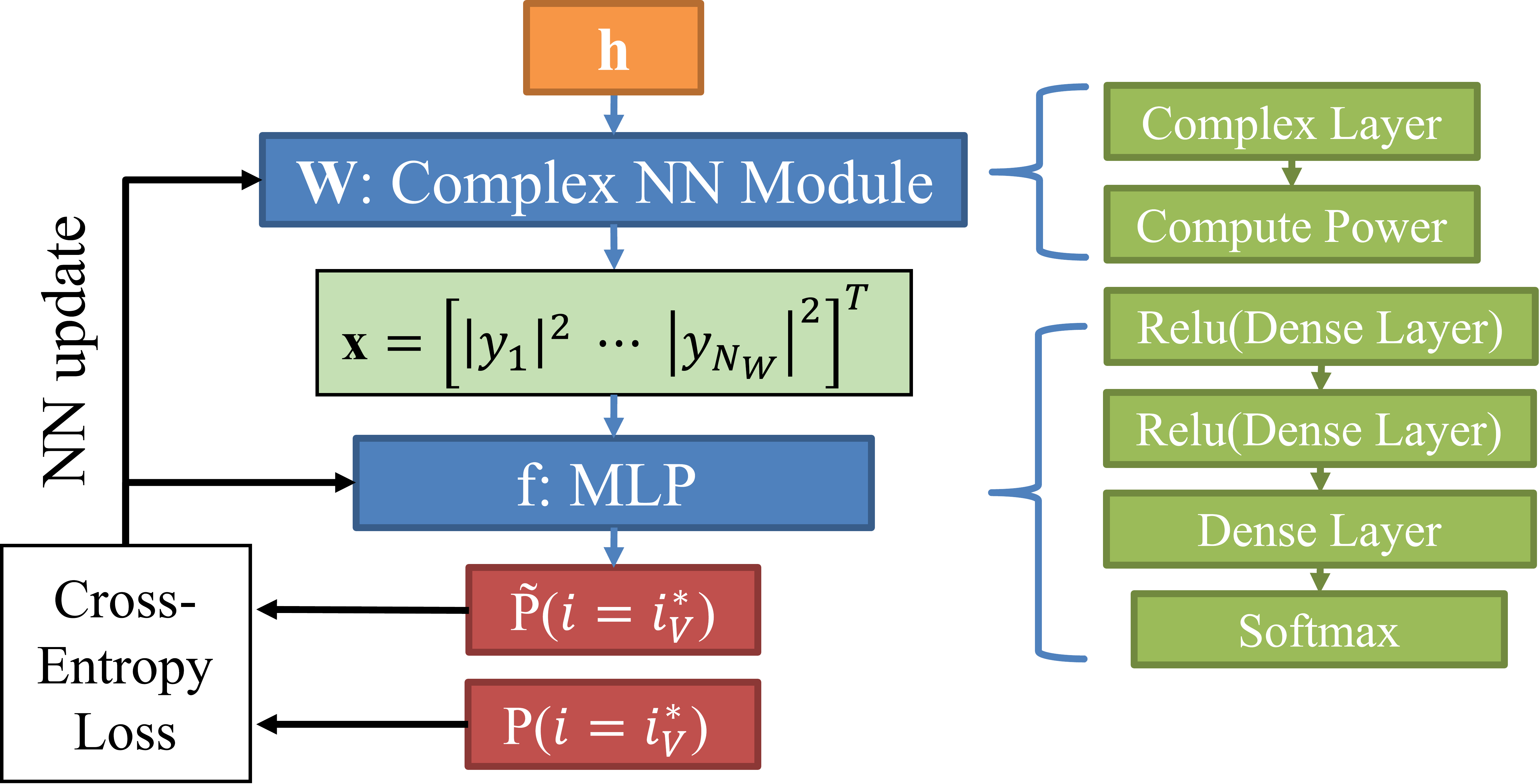}
  \caption{The architecture of the proposed NN, including the probing codebook $\mathbf{W}$ and the beam selection function $f$.}\label{figure:NN_architecture_training}
\end{figure}

The probing codebook $\mathbf{W}$ needs to be designed so that its beam sweeping measurements provide useful information regarding which narrow beam in $\mathbf{V}$ to select. A good probing codebook is site-specific and should capture particular characteristics of the propagation environment. The beam selection function $f$ is also optimized for that particular \ac{BS} and needs to be designed so that it picks narrow beams from $\mathbf{V}$ which tend to maximize the average \ac{SNR}. Since the probing codebook $\mathbf{W}$ and the beam selection function $f$ are interdependent, they are parameterized with different \ac{NN} modules -- a complex-\ac{NN} module and a \ac{MLP} classifier -- and jointly trained in an end-to-end fashion. The overall architecture is illustrated in Fig. \ref{figure:NN_architecture_training}.

\subsubsection{The trainable probing codebook}\label{section:complex_NN_module}
Beam sweeping using the probing codebook is modeled with a complex-\ac{NN} module that computes the complex received \ac{BF} signals and their power. The input to this \ac{NN} module is the channel vector $\mathbf{h}$. The complex layer in the complex-\ac{NN} module implements the complex arithmetic of analog \ac{BF} using real arithmetic. When parameterizing a $N_{\mathbf{W}}$-beam codebook, the trainable weights of the complex layer are elements of $\Theta \in \mathbbm{R}^{N_t \times N_{\mathbf{W}}}$, which are the phase shift values applied to each antenna element. The complex \ac{BF} weights $\mathbf{W} \in \mathbbm{C}^{N_t \times N_{\mathbf{W}}}$ can then be computed as 
\begin{equation}
    \mathbf{W} = \frac{1}{\sqrt{N_t}}(\cos{\Theta}+j\cdot\sin{\Theta}).
\end{equation}
The complex matrix multiplication
\begin{equation}
    \mathbf{z} = \mathbf{W}^H \mathbf{h}
\end{equation}
can be expressed as a real matrix multiplication
\begin{equation}
\begin{bmatrix}
\mathbf{z}^{real} \\
\mathbf{z}^{imag}
\end{bmatrix}
=
\begin{bmatrix}
\mathbf{W}^{real} & -\mathbf{W}^{imag}\\
\mathbf{W}^{imag} & \mathbf{W}^{real}
\end{bmatrix}^T
\begin{bmatrix}
\mathbf{h}^{real} \\
\mathbf{h}^{imag}
\end{bmatrix}
,
\end{equation}
where $\mathbf{h} \in \mathbbm{C}^{N_t \times 1}$ is the channel vector, $\mathbf{z} \in \mathbbm{C}^{N_\mathbf{W} \times 1}$ is the \ac{BF} output, and we express $\mathbf{z},\mathbf{W}$ and $\mathbf{h}$ in terms of their real and imaginary parts.
The \ac{BF} signal power can then be computed as 
\begin{equation}
    |\mathbf{z}|^2 = \left[(\mathbf{z}_1^{real})^2+(\mathbf{z}_1^{imag})^2,\cdots,(\mathbf{z}_{N_{\mathbf{W}}}^{real})^2+(\mathbf{z}_{N_{\mathbf{W}}}^{imag})^2\right]^T.
\end{equation}
While $|\mathbf{z}|^2$ is not complex differentiable with respect to $\mathbf{z}$, backpropagation can be enabled by treating the real and imaginary parts of $\mathbf{z}$ independently and compute $\frac{\partial |\mathbf{z}|^2}{\partial \mathbf{z}^{real}}$ and $\frac{\partial |\mathbf{z}|^2}{\partial \mathbf{z}^{imag}}$. The phase-shift values $\Theta$ can then be updated using the chain rule and backpropagation. 
Since only the dimension of $\Theta$ needs to be specified when initializing the \ac{NN}, the complex-\ac{NN} module only needs to know the number of antenna elements and not the exact array geometry. The array-geometry information is embedded in the input channel vectors so that the complex-\ac{NN} module can automatically learn the optimal phase-shift values to apply at each antenna element. This allows the architecture to be flexibly adopted by \acp{BS} with different antenna arrays.

The complex-\ac{NN} module computes the received signal power in (\ref{eq:bf_signal_power}). Since updates are made to the phase-shift values $\Theta$ during training, this architecture enforces the constant-modulus constraint of phase-shifter-only analog \ac{BF}. Note that $\Theta$ can be extracted from the \ac{NN} module at any time and be implemented as an analog codebook. After training, the complex-\ac{NN} module can be discarded so that (\ref{eq:bf_signal_power}) can be computed using an actual \ac{RF} chain and with an analog \ac{BF} codebook derived from $\Theta$. 

\subsubsection{The MLP beam selection function}
The beam selection function $f$ is modeled using an \ac{MLP} classifier, which is a feedforward fully connected \ac{NN} with non-linear activation functions.
The input to the \ac{MLP} is the power of the complex received signals of all beams in $\mathbf{W}$, which is calculated using the complex-\ac{NN} module during training or through beam sweeping during deployment. 
The \ac{MLP} consists of several hidden layers before the output layer to increase its approximation power. The output of an \ac{MLP} with 1 hidden layer can be written as
\begin{equation}\label{eq:mlp_1layer}
g(\mathbf{x}) = b_1 + A_1 \sigma(b_0 + A_0\mathbf{x}), 
\end{equation}
where the input feature vector is $\mathbf{x}$, the output vector is $g(\mathbf{x})$, the biases and weights of the hidden layer are $b_0$ and $A_0$, the biases and weights of the output layer are $b_1$ and $A_1$, and the non-linear activation function of the hidden layer is $\sigma$. The biases and weights models a trainable affine transformation, while the non-linear activation function allows the \ac{MLP} to approximate a wide range of non-linear functions. The biases and weights of the \ac{NN} can be updated through backpropagation to optimize some given objective function. 
One obvious way to optimize $f$ is to design it to predict the optimal narrow beam $\mathbf{v}^*$ which achieves the highest \ac{SNR} with the current channel. Hence, the final softmax layer of the \ac{MLP} outputs the predicted posterior probability distribution of each narrow beam in $\mathbf{V}$ being the optimal beam.
The \ac{MLP} is a powerful function approximator and can produce good estimates of posterior class probabilities. The \ac{BS} can select the narrow beam with the highest predicted posterior probability.
To increase the beam alignment robustness, the \ac{BS} can also use the output of the \ac{MLP} to reduce the search space and sweep the top-$k$ narrow beams with the highest predicted posterior probabilities. 

The complex-\ac{NN} architecture can be used to optimize a wide range of objective functions since it essentially implements analog \ac{BF} while allowing gradient descent updates that respect the phase-shifter-only constraints. For instance, it is used to directly minimize the the mean squared error (MSE) between the gain of the strongest beam in the codebook and the equal gain combining (EGC) gain in \cite{alrabeiah2020NNcodebook_arxiv}. It is also shown to be robust against hardware impairment \cite{alrabeiah2020NNcodebook_arxiv}. In order to select an optimal beam from a given narrow-beam codebook, the probing codebook should provide useful information to the beam selection function based on the entire environment. Hence the \ac{MLP} beam selection function is stacked after the complex-\ac{NN} module and the entire \ac{NN} is trained in an end-to-end fashion instead of directly optimizing the \ac{BF} gain of the probing beams.
The cross-entropy between the predicted optimal-beam distribution and the true optimal-beam distribution is used as the loss function. The partial derivative of the loss function with respect to the \ac{MLP} biases and weights as well as $\Theta$ in the complex-\ac{NN} module can be computed so that the \ac{MLP} and the complex-\ac{NN} module can be updated during training. The probing codebook is optimized implicitly to assist the downstream beam selection function. Interestingly, it still learns to capture particular characteristics of the propagation environment, as will be discussed in Section \ref{section:probing_codebook_pattern}.

\subsection{Practicality of the proposed method in 5G}
The proposed beam alignment method requires an offline training phase and a deployment phase. During the training phase, the \ac{BS} optimizes the probing codebook and the beam selection function by learning from training data and updating the \ac{NN}. The training data consists of the channel vectors for a \ac{BS} and its potential \acp{UE}. 
Operators can obtain the channel vectors through ray-tracing simulations of the site prior to deployment. 
Alternatively, or for further refinement of the probing codebook, the \ac{BS} can begin with a default codebook, and then gradually develop a site-specific probing codebook through interaction with its \acp{UE}.  In a typical \ac{TDD} scenario, the \ac{BS} can directly estimate the \ac{UL} channel by receiving the \acp{SRS} transmitted by \acp{UE} and assume the \ac{DL} channel is same as the estimated \ac{UL} channel. If the \ac{DL} and \ac{UL} channel reciprocity does not exist, the \ac{UE} can estimate the \ac{DL} channel by receiving the \ac{SSB} and \ac{CSI-RS} transmitted by the \ac{BS} and then feed back the estimated channel.

\begin{figure}[!tbp]
  \centering
  \includegraphics[width=0.7\textwidth]{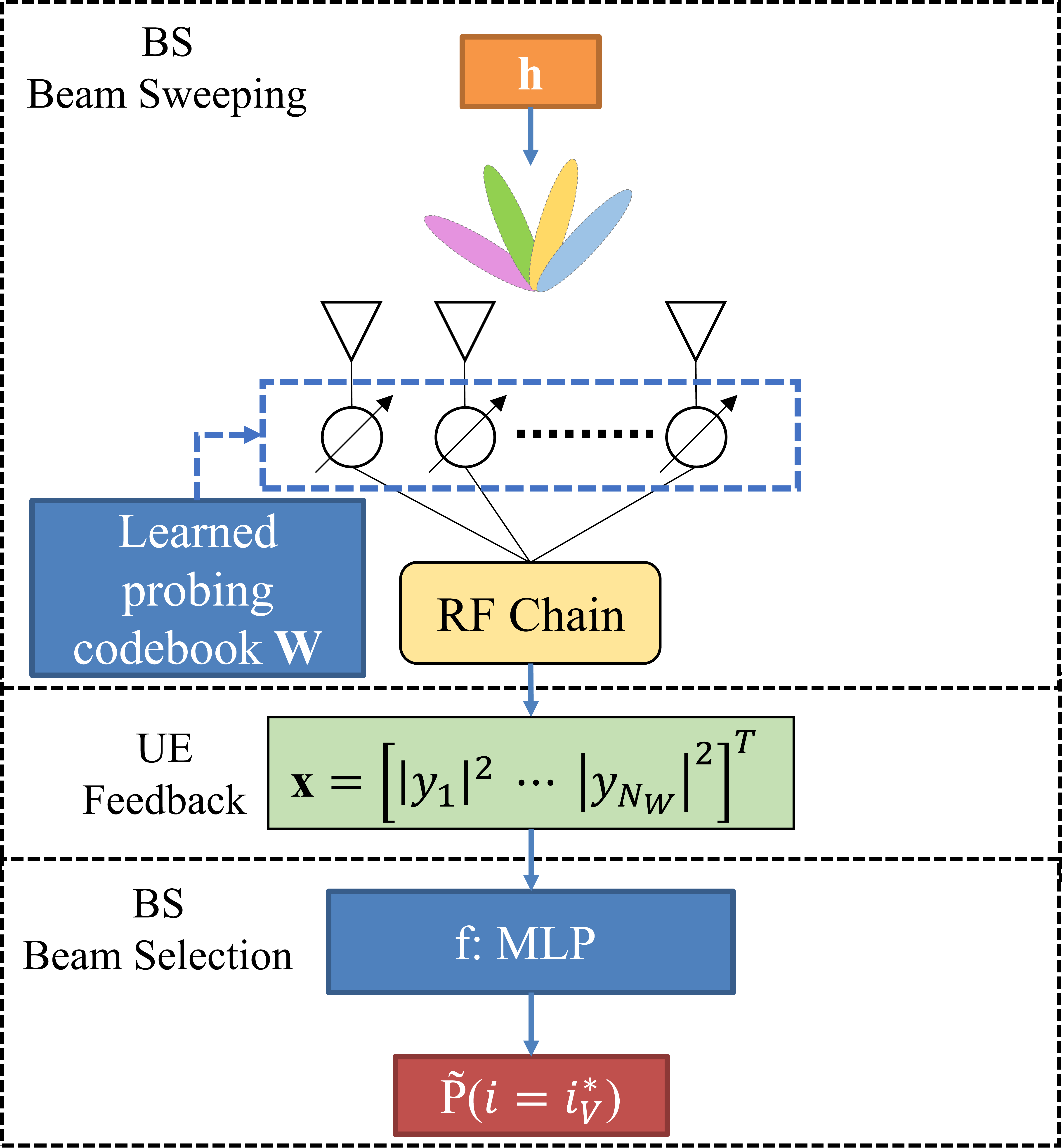}
  \caption{Proposed beam alignment architecture during the deployment phase.}\label{figure:NN_architecture_deployment}
\end{figure}

While the probing codebook is parameterized using a \ac{NN} during training, the complex-\ac{NN} module can be discarded in the deployment phase. As illustrated in Fig. \ref{figure:NN_architecture_deployment}, the \ac{BF} weights of the probing codebook can be extracted from the complex-\ac{NN} module and implemented as an analog beam codebook at the \ac{BS} after training. 
During the deployment phase, the \ac{BS} periodically sweeps the learned probing codebook by transmitting a sequence of \acp{SSB} using different probing beams. Each \ac{UE} measures all the \acp{SSB} and reports the received signal power to the \ac{BS}. The received signal power vector $\mathbf{x}$ is fed into the \ac{MLP} beam predictor at the \ac{BS}, which then selects the optimal narrow beam or a few candidate beams to try according to the predicted posterior probability distribution. If the \ac{BS} chooses to search the top-$k$ predicted narrow beams for additional robustness, it can do so by sweeping those beams using the aperiodic \acp{CSI-RS}, which can be independently configured for each \ac{UE}.
The proposed beam alignment method is adapted to characteristics of the propagation environment such as the distribution of \acp{UE} and the location of the scatterers. If the environment changes, the probing codebook as well as the \ac{MLP} beam predictor need to retrained. The \ac{NN} modules can be trained from scratch, or be initialized with the existing probing codebook and MLP weights and be refined using data from the new environment. The retraining can be triggered if the beam alignment performance is below a threshold. Since such macroscopic characteristics of the environment are expected to evolve slowly, the retraining should occur infrequently.

The proposed beam alignment method essentially consists of periodic beam sweeping by the \ac{BS} and beam measurement and reporting by the \ac{UE} during the deployment phase. This beam sweeping, measurement and reporting process precisely fits into the beam sweeping-based framework currently adopted in 5G, as discussed in Section \ref{section:related_work}. Instead of sweeping a general codebook using \acp{SSB} to cover the entire angular space, the proposed method sweeps a site-specific probing codebook that learns to strategically place beams in directions that can effectively capture characteristics of the environment and provide useful information to the downstream beam selector. Since the learned probing codebook replaces traditional full-coverage codebooks, its beams can be transmitted using ``always-on" \acp{SSB} and thus can also be used by unconnected \acp{UE} for cell discovery and \ac{IA}.
Instead of selecting the narrow beam with the highest reported power and ignoring the measurement of the rest of the codebook, the proposed method predicts good candidate beams using a \ac{NN} that intelligently utilizes the measurement of all probing beams. This is also supported in 5G since the \ac{BS} can request additional beam reports from \acp{UE} to obtain measurements of all probing beams.  
Overall, the proposed method can be directly adopted in 5G and does not require any modification to the existing 5G standard.


\subsection{Baselines and Metrics}
The proposed beam alignment method selects the optimal narrow beam based on measurements of a probing codebook. It is analogous to a hierarchical beam search where the optimal narrow child beam is determined based on measurements of wider parent beams. In a traditional hierarchical beam search, the \ac{BS} needs to sweep all child beams of the best parent beam at each layer of the codebook. The parent beam needs to have a wider beam width and should contain the coverage areas of its child narrower beams. Four baselines of comparison are considered, including 2 hierarchical beam searches, an exhaustive beam search and a genie. In all baselines, the \ac{BS} has the same narrow-beam codebook $\mathbf{V} \in \mathbbm{C}^{N_t \times N_{\mathbf{V}}}$ with $N_{\mathbf{V}}$ beams from which it needs to select a beam for the data or control channel. When comparing the performance of different beam alignment methods, the beam alignment accuracy and the \ac{SNR} are two key metrics. The beam alignment accuracy is the probability or relative frequency that the \ac{BS} selects the optimal narrow beam from the codebook $\mathbf{V}$. The \ac{SNR} is calculated as in (\ref{eq:snr}).

\textbf{2-Tier Hierarchical Beam Search} The 2-tier hierarchical beam search uses a wide-beam codebook with $N_{\mathbf{W}}$ beams and a narrow-beam codebook with $N_{\mathbf{V}} \gg N_{\mathbf{W}}$ beams, both covering the same angular space. Each narrow beam is the child beam of one of the wide beams. The coverage area of each wide beam contains the coverage areas of all of its children beams. The \ac{BS} first sweeps the $N_{\mathbf{W}}$ wide beams then sweeps the children beams of the best wide beam. The final selected beam is the best child beam of the best wide beam. The wide-beam codebook is analogous to the proposed probing codebook in that they both provide rough information about the channel for beam selection.

\textbf{Binary Hierarchical Beam Search} The binary hierarchical beam search is a generalized version of the 2-tier beam search. It performs a binary tree search on the narrow beam codebook $\mathbf{V}$. Starting with a search space equal to the entire angular space, the \ac{BS} repeatedly splits the search space into two partitions and sweeps two wide beams each covering one of the partitions until reaching one of the narrow beams in the final codebook $\mathbf{V}$. With $N_{\mathbf{V}}$ narrow beams, each beam search consists of $\log_{2}N_{\mathbf{V}}$ layers. With more hierarchical search layers, the binary beam search is more susceptible to search errors compared to the 2-tier search. If a sub-optimal wide beam is chosen in any of the upper layers due to noise or imperfect wide-beam patterns, the error will propagate forward and affect the selected narrow beam in $\mathbf{V}$.

\textbf{Exhaustive Beam Search} The \ac{BS} exhaustively sweeps the narrow beam codebook $\mathbf{V}$ with $N_{\mathbf{V}}$ beams and selects the beam with the highest received power. Compared to a hierarchical beam search, the exhaustive search directly measures the narrow-beam codebook instead of some wide-beam intermediate codebooks. The best beam in the narrow-beam codebook has larger directionality gain compared to the wide beams used in the hierarchical methods. As a result, the exhaustive search is less susceptible to noise in the beam measurements. 

\textbf{Genie (Upper Bound)} The \ac{BS} has knowledge of the true \ac{BF} gain of each narrow beam in the codebook $\mathbf{V}$ and always selects the best beam. While the hierarchical searches and the exhaustive search are all susceptible to search errors caused by noise in the received \ac{BF} signal, the genie method is not. If the receive noise power is zero, the genie is equivalent to the exhaustive beam search; else it is strictly better than exhaustive search. Given a narrow-beam codebook $\mathbf{V}$, the genie method achieves a perfect beam alignment accuracy of 100\% and provides a performance upper bound.

The hierarchical beam search approaches require wide beams that contain the coverage area of their children narrower beams. Multiple works have studied hierarchical codebook designs. Synthesizing wide beams often requires multiple \ac{RF} chains or antenna activation, such as in \cite{alkhateeb2014hybridBF}, \cite{noh2017multi_res_codebook} and \cite{xiao2016hierarchical}. Since a single \ac{RF} chain and analog \ac{BF} only is assumed in this work, the \ac{AMCF} algorithm proposed in \cite{qi2020hierarchical} is used to generate the wide beams in the hierarchical codebooks. 

\section{Dataset}\label{section:dataset}
Realistic and accurate data is essential to learning good \ac{NN} models. Ray tracing is able to achieve high accuracy and maintain spatial consistency when modeling \ac{mmWave} channels. A state-of-the-art commercial-grade ray-tracing software called Wireless InSite \cite{Remcom01} is used to generate the channel data. The ray-tracing software simulates rays emitting from the transmitter at all directions in the angular space and computes their interaction with the environment along their paths before reaching the receiver, including scattering, reflection and blockage. An environment needs to be constructed in the ray-tracing software, specifying the terrain, the scatterers and their dielectric properties.

Four different ray-tracing scenarios are considered to capture a wide range of propagation environments for \ac{mmWave}: a dense urban outdoor area, an urban street, an indoor conference room with hallways and an urban street with severe blockage and reflections. The ray-tracing scenarios include both \ac{LOS} and \ac{NLOS} \acp{UE} and cover two different \ac{mmWave} carrier frequencies: 28 GHz and 60 GHz. The ray-tracing simulation parameters are summarized in Table \ref{table:data_params}.

\begin{figure}
\centering
\includegraphics[width=0.6\linewidth]{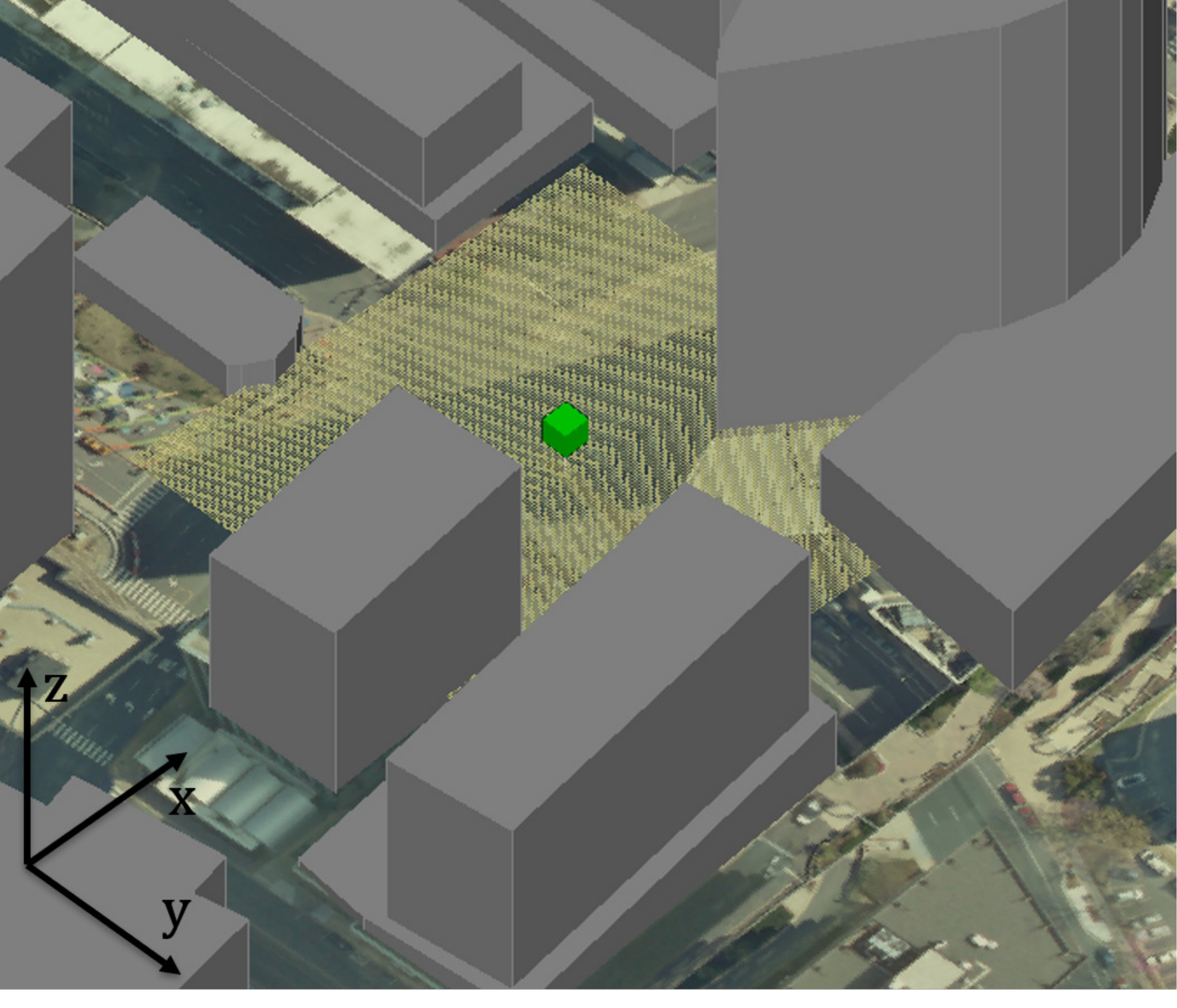}
\caption{3-D view of the Rosslyn environment. The green block represents the BS. Points on the yellow grid represent the UEs.}\label{figure:rosslyn_3D}
\end{figure}

\textbf{Rosslyn Experiment} The Rosslyn dataset captures an outdoor dense urban environment located in downtown Rosslyn, Virginia, USA. It was created with our own experiments and was published in \cite{heng2021MLbeamalignment}. A 3-D render of the environment is shown in Fig. \ref{figure:rosslyn_3D}. The Rosslyn environment has multiple buildings surrounding an intersection. A \ac{BS} is placed at the center of the intersection, elevated by 10 meters above the ground. A total of 73,884 \ac{UE} positions are placed uniformly 0.35 meter apart and 2 meters above the terrain surface. The entire simulated area is around 90 meters $\times$ 90 meters. The Rosslyn environment consists of mostly \ac{LOS} \acp{UE} and uses a carrier frequency of 28 GHz. 

\textbf{DeepMIMO O1\_28 Experiment} The DeepMIMO O1\_28 dataset captures an outdoor street environment and is available in the public DeepMIMO dataset \cite{alkhateeb2019deepmimo}. A portion of the original dataset corresponding to \ac{BS} \#3 and \acp{UE} in row \#800 to row \#1200 is selected. The environment consists of a street with buildings on both sides. The \ac{BS} is placed on one side of the street with an elevation of 6 meters. A total of 72,581 \ac{UE} positions are placed uniformly on the street 20 centimeters apart. The environment consists of mostly \ac{LOS} \acp{UE} and uses a carrier frequency of 28 GHz.

\begin{figure}
\centering
\includegraphics[width=0.6\linewidth]{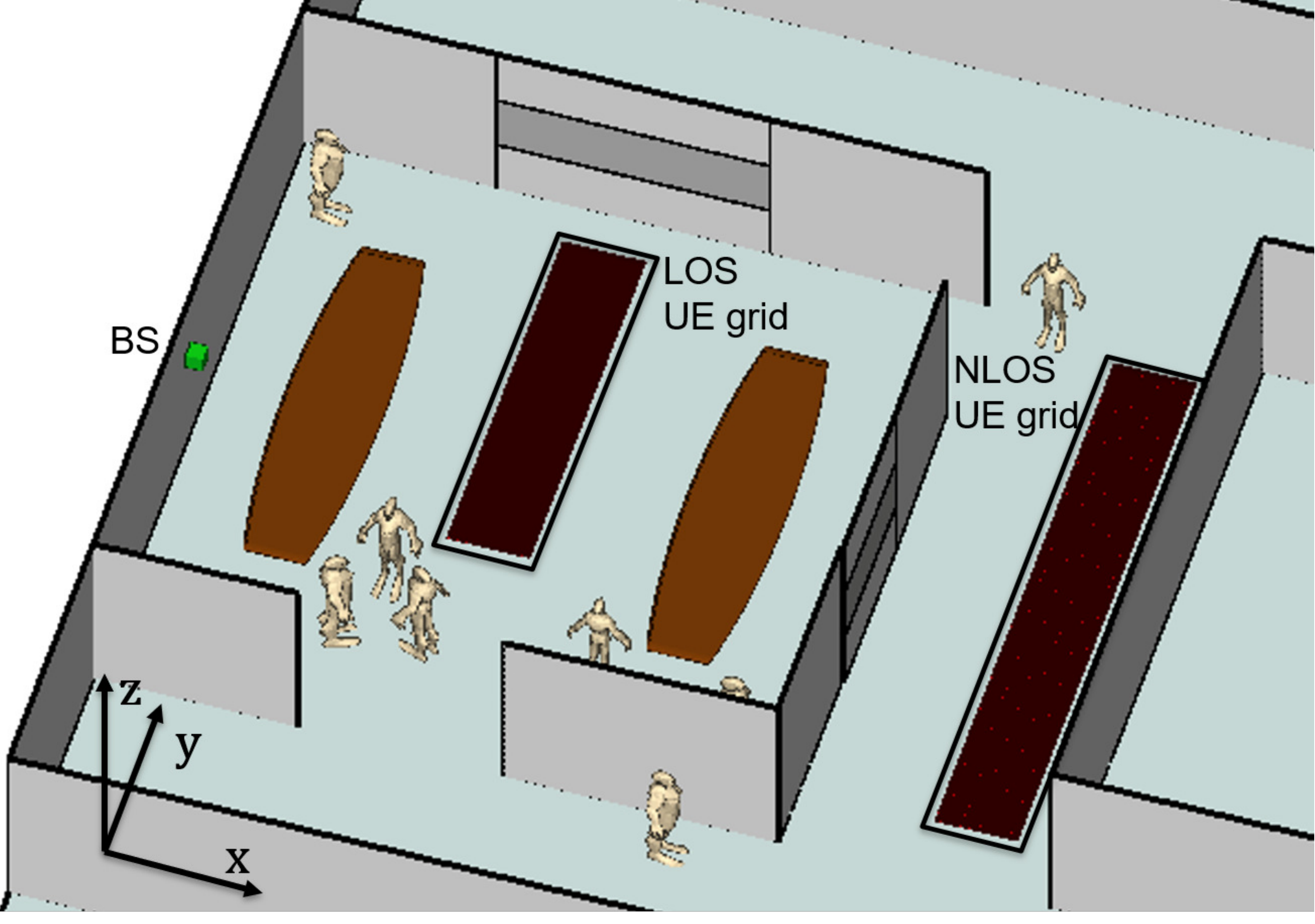}
\caption{3-D view of the DeepMIMO I3 environment. Adapted: \cite{deepmimo_scenario_webpage}}\label{figure:deepmimo_I3_3D}
\end{figure}

\textbf{DeepMIMO I3 Experiment} The DeepMIMO I3 dataset models an indoor conference room and its hallways and is available in the public DeepMIMO dataset \cite{alkhateeb2019deepmimo}. A 3-D view of the environment is shown in Fig. \ref{figure:deepmimo_I3_3D}. A \ac{BS} is placed 2 meters high on the wall inside the conference room. A total of 118,959 \ac{UE} positions are placed inside two grids: one \ac{LOS} grid inside the conference room and one \ac{NLOS} grid in the hallway. The carrier frequency is 60 GHz.

\begin{figure}
\centering
\includegraphics[width=0.6\linewidth]{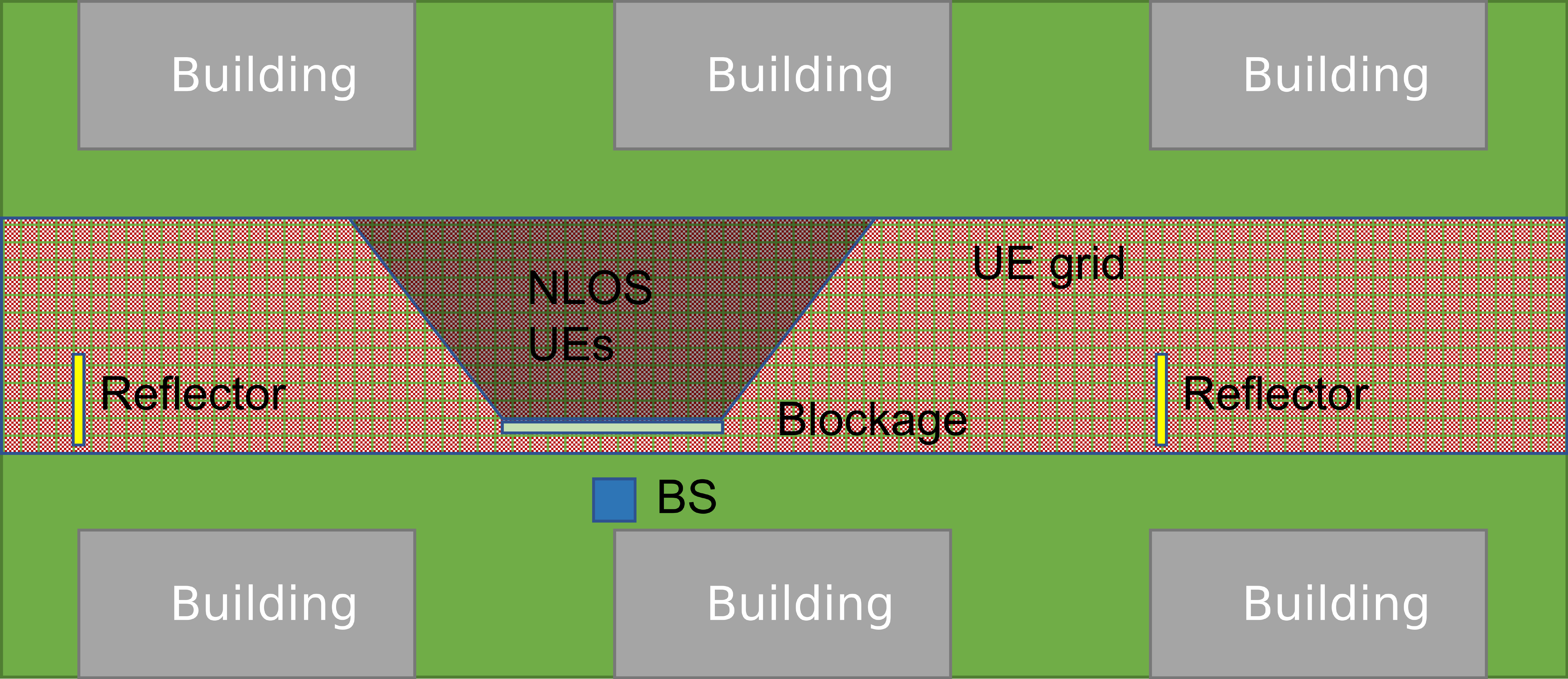}
\caption{2-D illustration of the DeepMIMO O1\_28B environment. Dimensions are not to scale.}\label{figure:deepmimo_028b_2D}
\end{figure}

\textbf{DeepMIMO O1\_28B Experiment} The DeepMIMO O1\_28B outdoor street environment is similar to the O1\_28 scenario \cite{alkhateeb2019deepmimo} but with severe blockage and reflections. A 2-D illustration of the environment is shown in Fig. \ref{figure:deepmimo_028b_2D}. A 24-meter-wide metal screen is placed in front of the \ac{BS} and two reflectors are placed on both sides. A total of 497,931 \ac{UE} positions are placed uniformly on the street 20 centimeters apart. This environment includes both \ac{LOS} and \ac{NLOS} \ac{UE} and uses a carrier frequency of 28 GHz.

\section{Evaluation}\label{section:evaluation}
Accurate ray-tracing channel data is used in our experiments as described in Section \ref{section:dataset}. In all experiments, 60\% of the data is used for training, 20\% of the data is used for validation and the remaining 20\% is used for testing. The training dataset is used to optimize the \ac{NN} weights. The hyperparameters of the \ac{NN} are tuned by performing a grid search over a set of predefined values and comparing their performance on the validation set. The test set is used to evaluate the final performance of each beam alignment method. The \ac{MLP} module in the proposed NN architecture has 2 hidden layers with rectified linear unit (ReLU) activation. The \ac{NN} is trained for 200 epochs using the Adam optimizer \cite{Kingma2015AdamAM}. The training and validation loss after each training epoch is examined to ensure that the \ac{NN} has converged. A complex \ac{AWGN} is assumed. The noise power in the received \ac{BF} signal in (\ref{eq:bf_signal}) is -81 dBm unless otherwise specified. The simulation parameters are summarized in Table \ref{table:data_params}. To make training more stable and efficient, the channel vectors are normalized by the maximum magnitude of the elements in the dataset: $\max_{\mathbf{h}  \in \mathcal{H}}|\mathbf{h}_{i,j}|$. Similar normalization techniques are adopted in \cite{alkhateeb2018deep} and \cite{alrabeiah2020NNcodebook_arxiv}. The noise is also scaled appropriately according to the normalization factor. Since the normalization factor is a predetermined constant that only depends on the underlying environment, it should not affect the practicality of the proposed method. The final narrow beam codebook $\mathbf{V}$ is a 128-beam \ac{DFT} codebook. 

\begin{table*}
\centering
  \caption{Simulation Parameters}\label{table:data_params}
    \begin{tabular}{| c | c |}
    \hline
    BS Antenna & $64\times1$ ULA\\ \hline
    UE Antenna & Single\\ \hline
    Narrow beam codebook size $N_{\mathbf{V}}$ & 128\\ \hline
    Carrier Frequency & \makecell{Rosslyn, DeepMIMO O1\_28, O1\_28B: 28 GHz\\ DeepMIMO I3: 60 GHz}\\ \hline
    Bandwidth ($B$) & 100 MHz\\ \hline
    Transmit Power ($P_{T}$) & \makecell{Rosslyn, DeepMIMO O1\_28, I3: 10 dBm\\ DeepMIMO O1\_28B: 20 dBm}\\ \hline
    Noise power spectral density (PSD) & -161 dBm / Hz \\ \hline
    Number of Rays & 25 \\ \hline
    \end{tabular}
\end{table*}

\subsection{Can the proposed beam alignment method achieve good accuracy?}
\begin{figure*}%
\centering
\subfloat[Rosslyn]{\includegraphics[width=0.5\columnwidth,trim=0.3cm 0.3cm 1.2cm 1.3cm,clip]{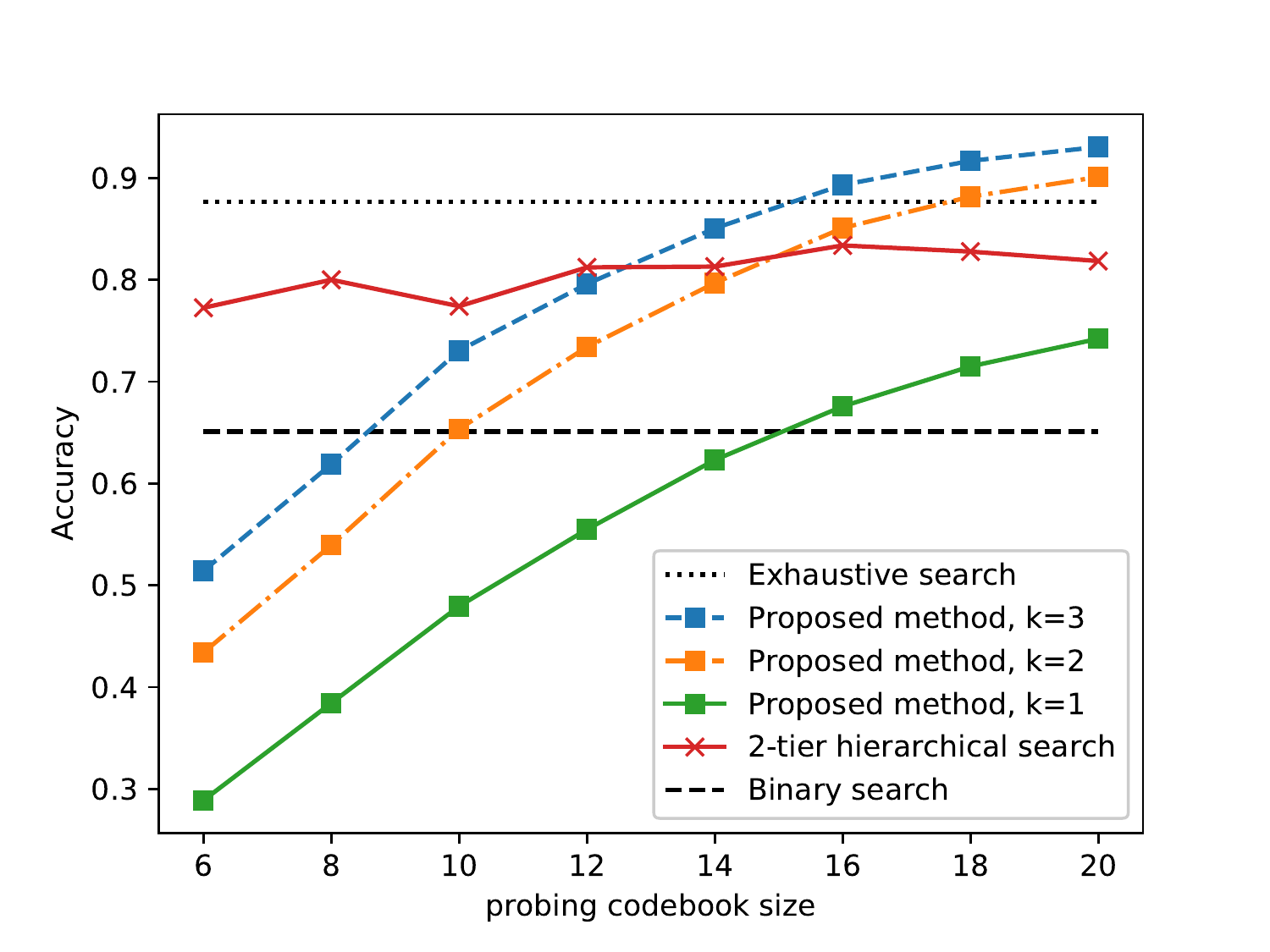}\label{figure:acc_vs_baselines_rosslyn}}
\subfloat[DeepMIMO O1\_28]{\includegraphics[width=0.5\columnwidth,trim=0.3cm 0.3cm 1.2cm 1.3cm,clip]{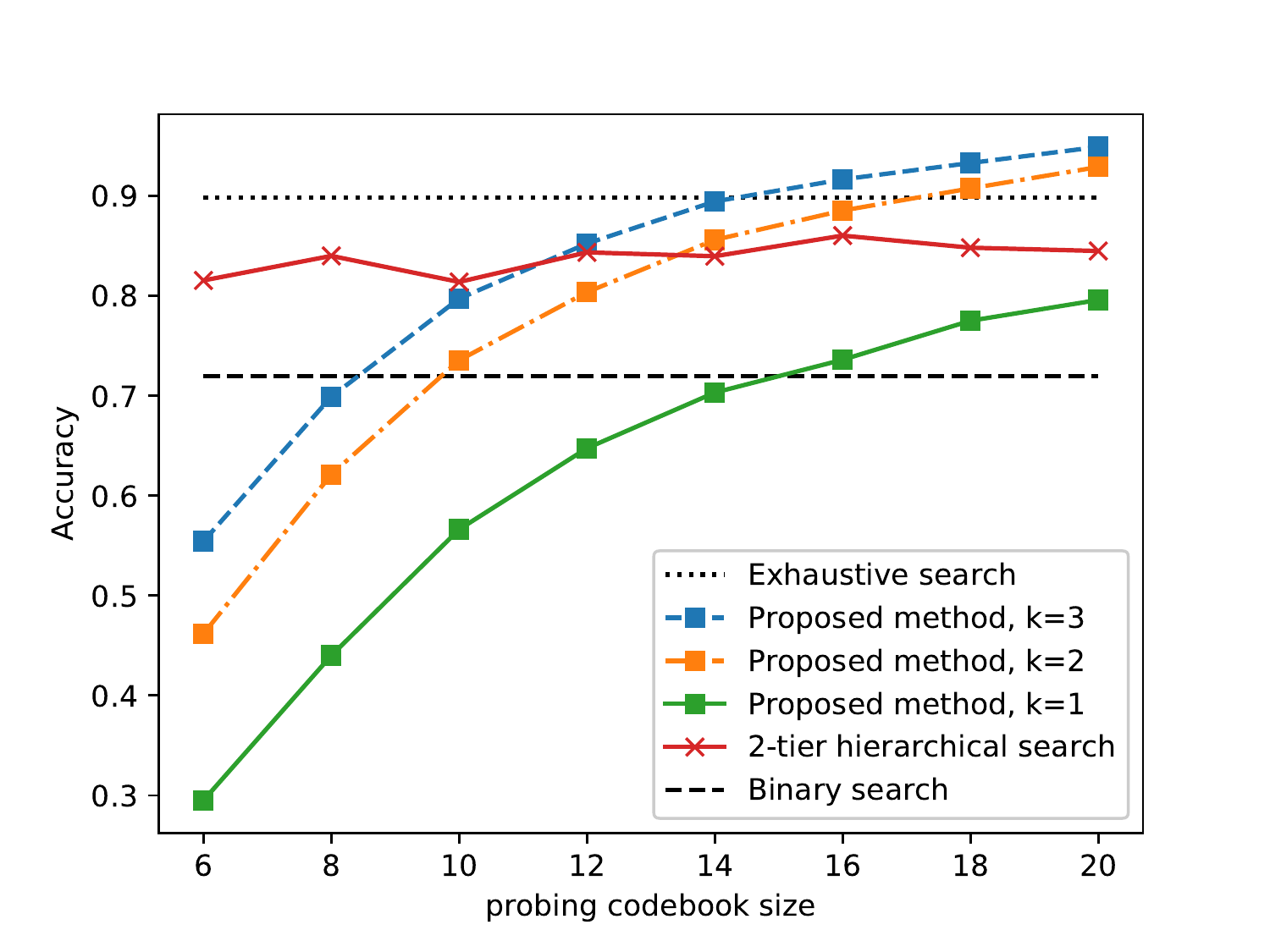}\label{figure:acc_vs_baselines_O1_28}}
\hfill
\subfloat[DeepMIMO I3]{\includegraphics[width=0.5\columnwidth,trim=0.3cm 0.3cm 1.2cm 1.3cm,clip]{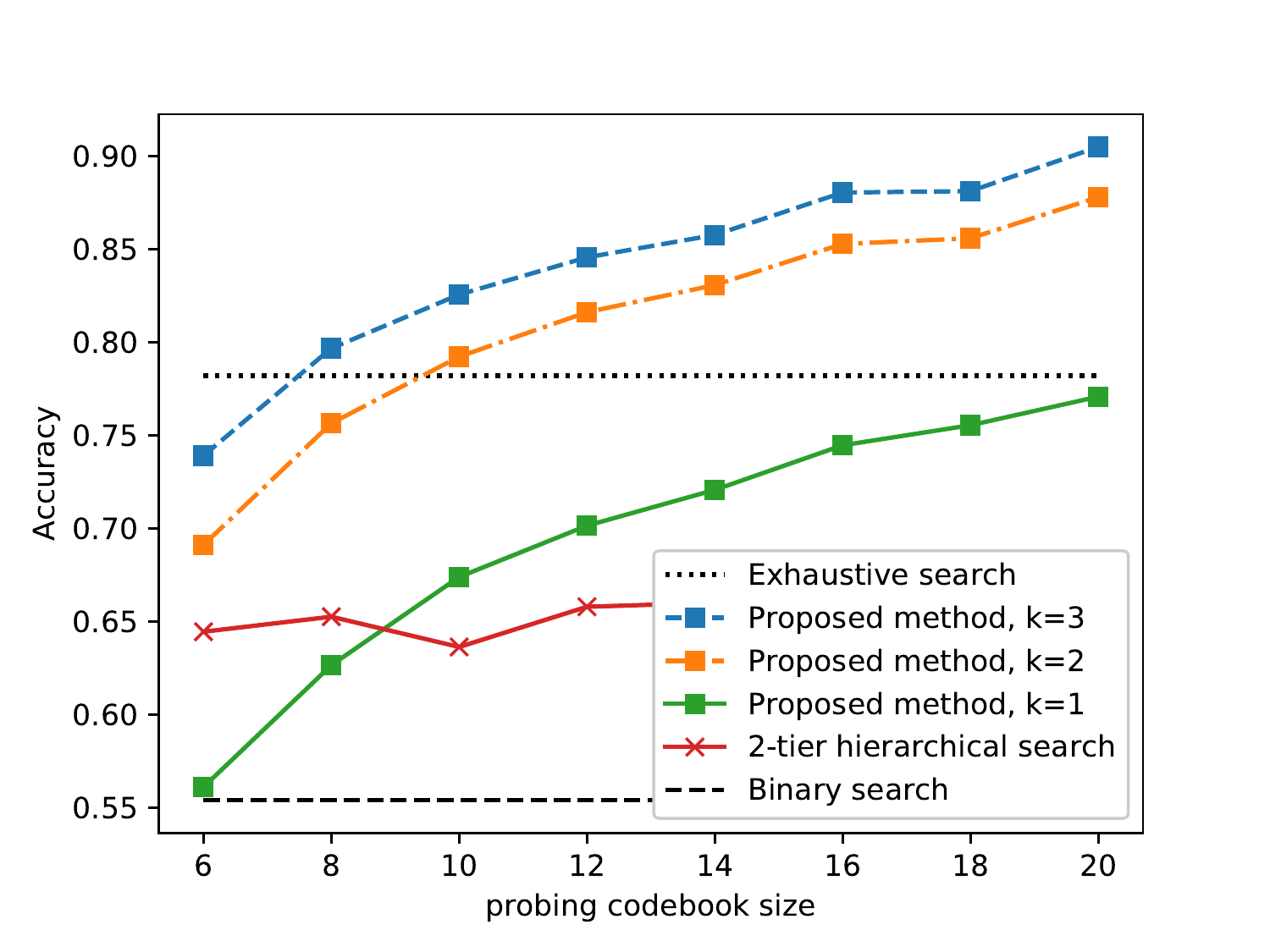}\label{figure:acc_vs_baselines_I3}}
\subfloat[DeepMIMO O1\_28B]{\includegraphics[width=0.5\columnwidth,trim=0.3cm 0.3cm 1.2cm 1.3cm,clip]{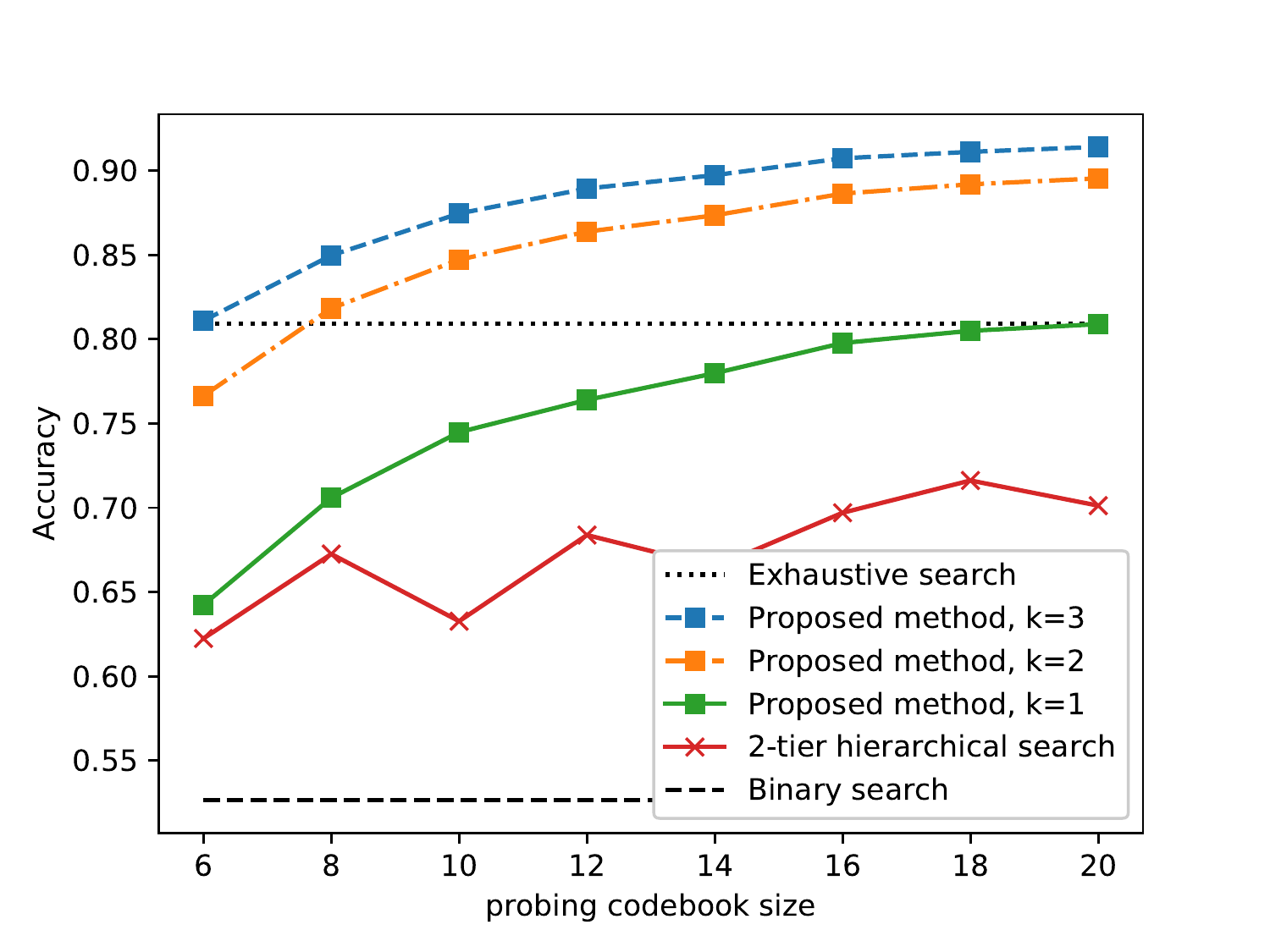}\label{figure:acc_vs_baselines_O1_28B}}
\caption{Beam alignment accuracy vs. probing codebook size.}\label{figure:acc_vs_baselines}
\end{figure*}

The accuracy of the proposed method and the baselines with increasing probing codebook size is shown in Fig. \ref{figure:acc_vs_baselines}. The genie always has a perfect accuracy of 1 and is not shown in the figure. The accuracy of the proposed method increases as the number of probing beams increases, which is expected since a larger probing codebook allows the \ac{BS} to obtain more information about the channel for a \ac{UE}. Among the traditional beam sweeping-based baselines, the exhaustive search performs the best and the binary search performs the worst. This is expected since a method with more layers in the hierarchical search structure is more vulnerable to noise in the received signal. 
Instead of directly choosing the predicted optimal beam, the accuracy of the proposed method can be improved significantly by searching a few additional candidate beams. In all 4 environments, the proposed method can achieve a beam alignment accuracy of at least 85\% with just 14 probing beams and searching an additional $k$ = 3 narrow beams, outperforming both the binary and the 2-tier hierarchical beam search baselines. 
In the 2 \ac{LOS} environments (Rosslyn and DeepMIMO O1\_28), the proposed method can beat the binary beam search with 10 probing beams and $k$ = 3. With 16 probing beams, it can even outperform the exhaustive search by sweeping an additional $k$ = 3 narrow beams. 
In the 2 environments with \ac{NLOS} \acp{UE} (DeepMIMO I3 and O1\_28B), traditional beam sweeping-based baselines perform considerably worse compared to in the \ac{LOS} environments. 
The exhaustive search can only achieve an accuracy of around 80\% compared to around 90\% in the \ac{LOS} environments. The hierarchical beam searches suffer from even worse accuracy degradation, with the 2-tier hierarchical beam search achieving accuracies of around 65\% compared to around 80\% in the \ac{LOS} environments. Clearly, beam alignment for the \ac{NLOS} \acp{UE} is more challenging for traditional beam sweeping-based baselines.
On the other hand, the proposed method shines in these environments with \ac{NLOS} \acp{UE}. With just 8 probing beams and $k$ = 3, it outperforms any beam sweeping-based baseline, including the exhaustive search. With 12 probing beams and $k$ = 3, the proposed method can achieve an accuracy of over 84\% in the I3 environment and over 88\% in the O1\_28B environment. 
Overall, the proposed method can outperform the traditional beam sweeping-based baselines with a moderate probing codebook size, particularly in environments with \ac{NLOS} \acp{UE} which are usually challenging for beam alignment. 

\subsection{Can the proposed beam alignment method achieve good SNR?}
\begin{figure*}%
\centering
\subfloat[Rosslyn]{\includegraphics[width=0.5\columnwidth,trim=0.3cm 0.3cm 1.2cm 1.3cm,clip]{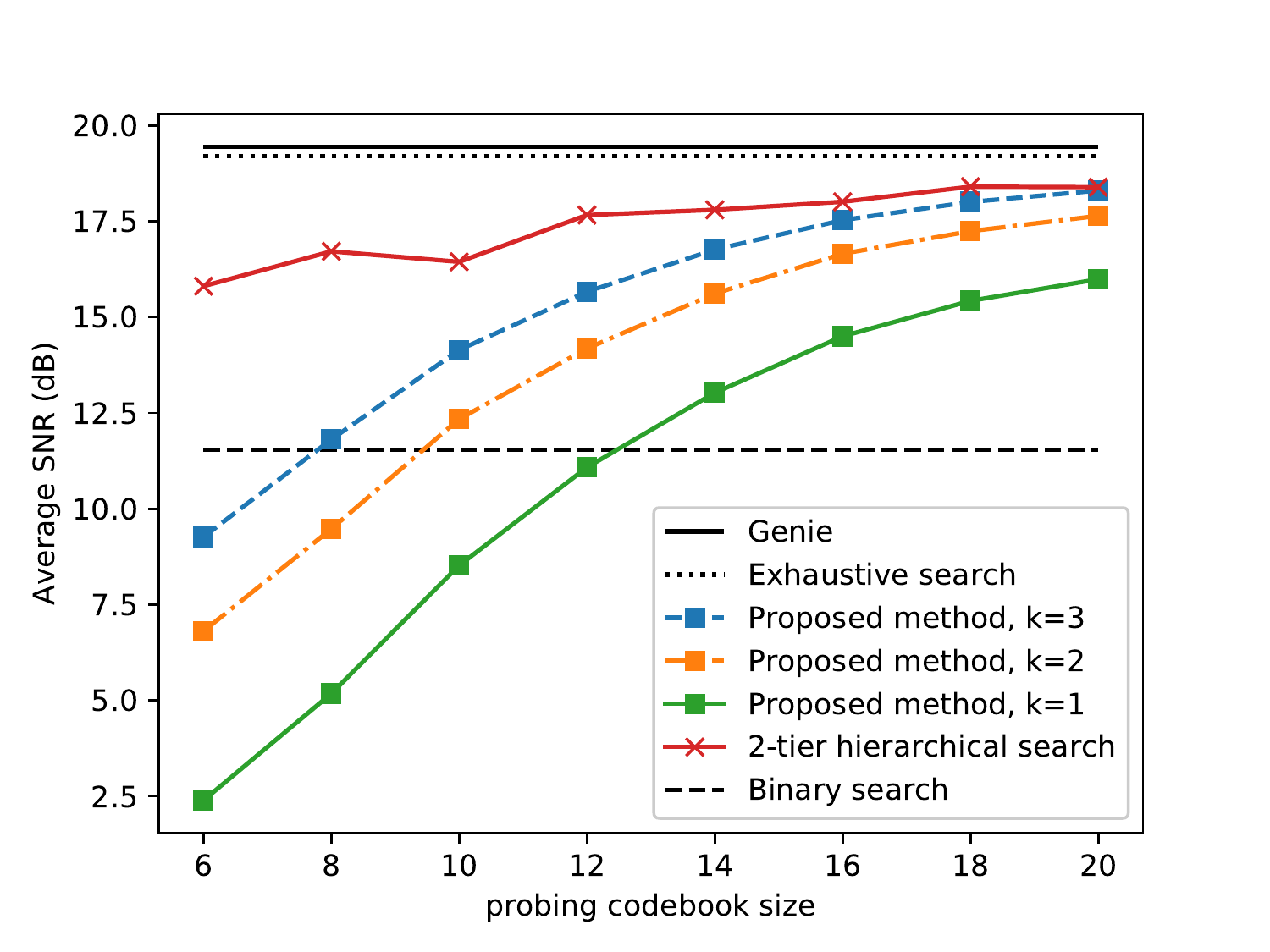}\label{figure:avg_snr_vs_baselines_rosslyn}}
\subfloat[DeepMIMO O1\_28]{\includegraphics[width=0.5\columnwidth,trim=0.3cm 0.3cm 1.2cm 1.3cm,clip]{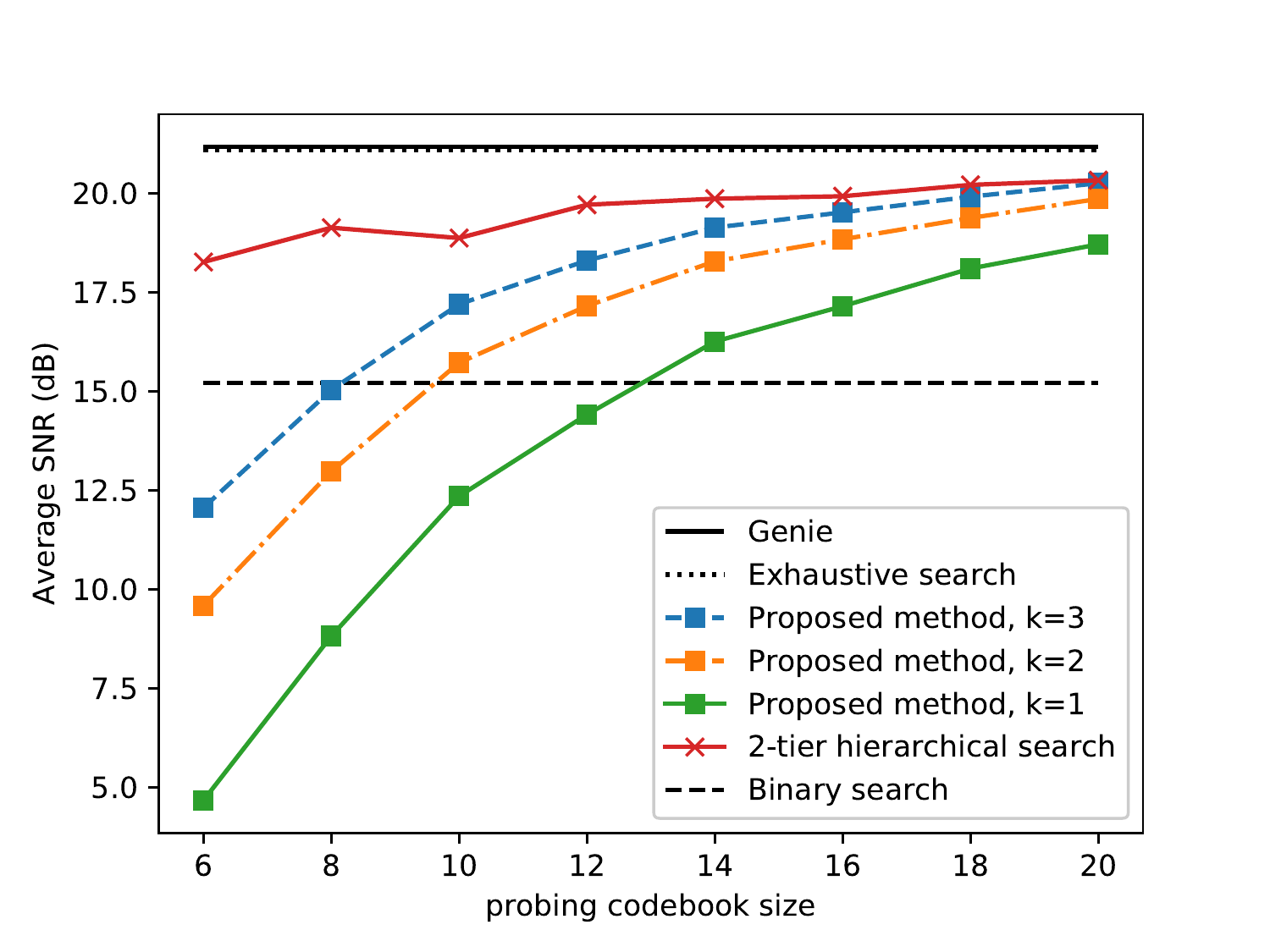}\label{figure:avg_snr_vs_baselines_O1_28}}
\hfill
\subfloat[DeepMIMO I3]{\includegraphics[width=0.5\columnwidth,trim=0.3cm 0.3cm 1.2cm 1.3cm,clip]{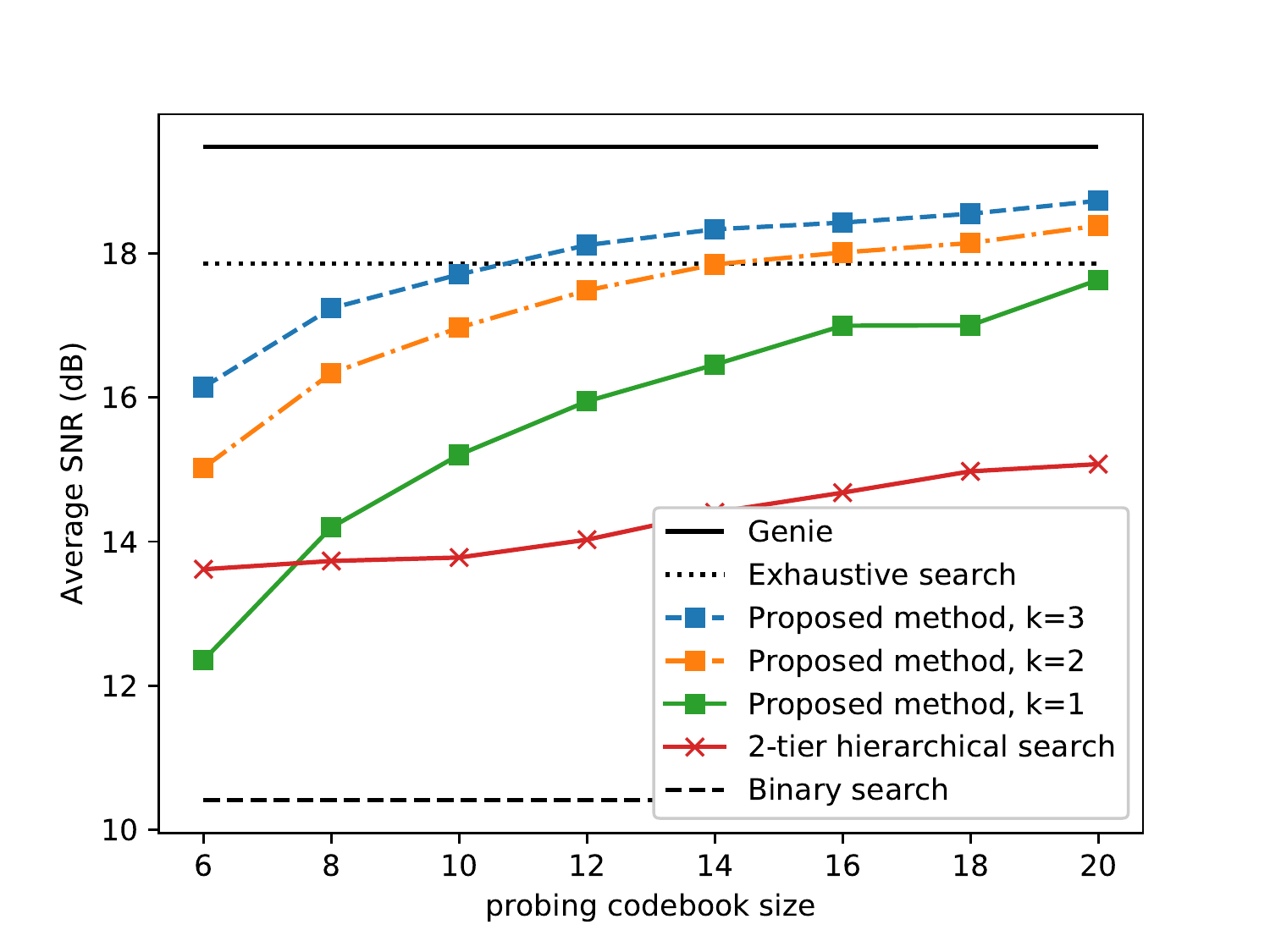}\label{figure:avg_snr_vs_baselines_I3}}
\subfloat[DeepMIMO O1\_28B]{\includegraphics[width=0.5\columnwidth,trim=0.3cm 0.3cm 1.2cm 1.3cm,clip]{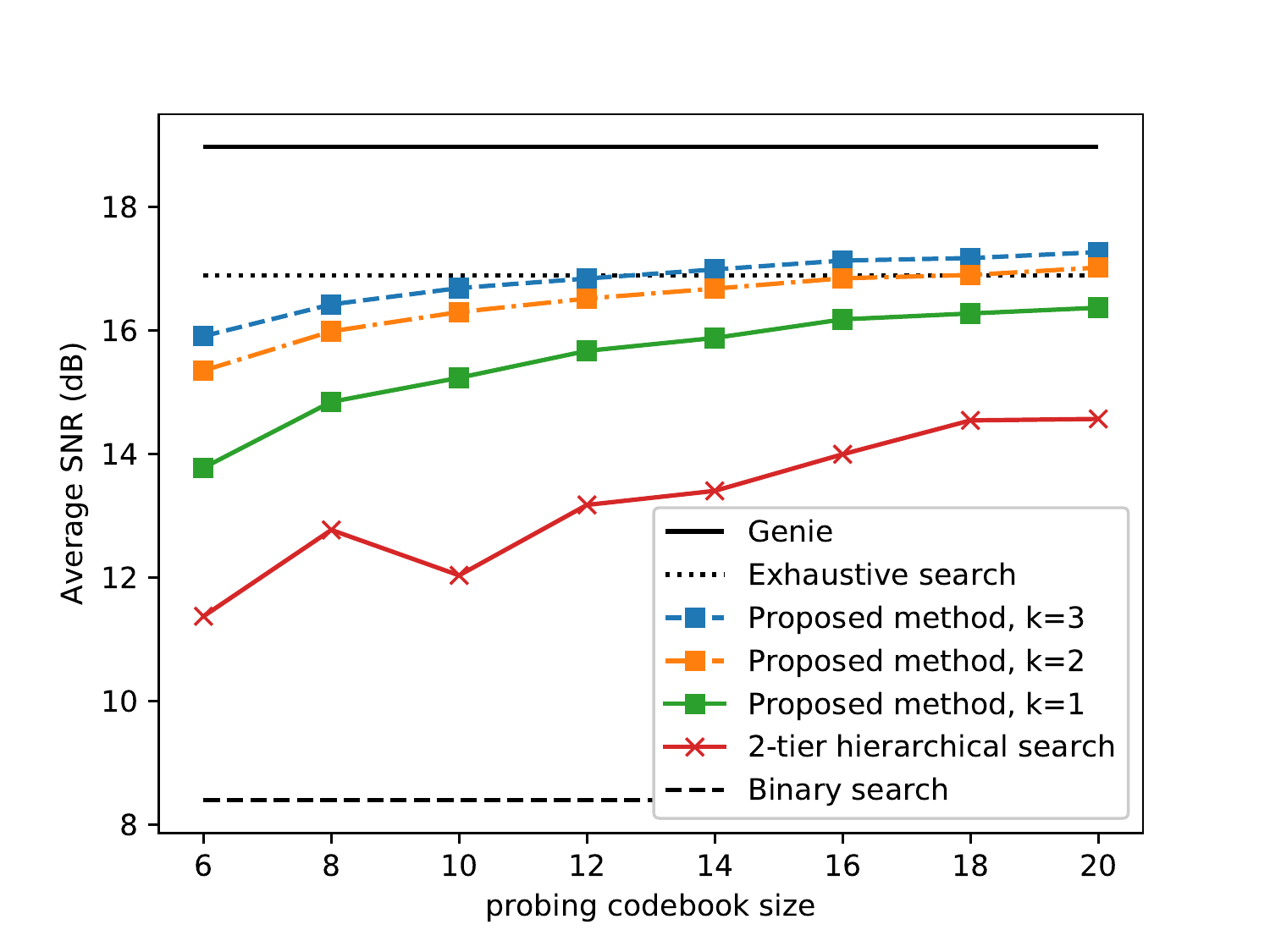}\label{figure:avg_snr_vs_baselines_O1_28B}}
\caption{Average SNR vs. probing codebook size.}\label{figure:avg_snr}
\end{figure*}

The beam alignment accuracy considers the probability of finding the optimal narrow beam. With a large, oversampled codebook, adjacent narrow beams may have similar \ac{BF} gains. As a result, operators may be more interested in the \ac{SNR} achieved after beam alignment.  
The average \ac{SNR} of the proposed method and the baselines with increasing probing codebook size is shown in Fig. \ref{figure:avg_snr}. 
In the \ac{LOS} environments (Rosslyn and DeepMIMO O1\_28), the proposed method outperforms the binary beam search with just 10 probing beams and $k$ = 2 but is worse than both the exhaustive and the 2-tier hierarchical methods. It is able to match the 2-tier hierarchical baseline in terms of average \ac{SNR} with 20 probing beams and $k$ = 3.
The exhaustive search achieves close-to-optimal average \ac{SNR} while its accuracy is just around 90\%.
Similar to the accuracy performance, the proposed method shines in the challenging \ac{NLOS} environments in terms of average \ac{SNR}. With just 8 probing beams and without additional narrow-beam sweeping, it outperforms both hierarchical search methods. With 12 probing beams and $k$ = 3, the proposed method can achieve similar if not better average \ac{SNR} compared to the exhaustive search. 
Overall, with 12 probing beams and $k$ = 3, the gap in the average SNR of the proposed method from the genie upper bound is 3.73 dB in the Rosslyn environment, 2.86 dB in the DeepMIMO O1\_28 environment, 1.36 dB in the I3 environment and 2.15 dB in the O1\_28B environment.

\subsection{Can the proposed NN learn meaningful probing codebooks?}\label{section:probing_codebook_pattern}
The learned probing codebook should provide meaningful and helpful information to the downstream \ac{MLP} beam predictor. In the proposed end-to-end training procedure, the complex-\ac{NN} probing codebook and the \ac{MLP} beam predictor are jointly trained. 
Intuitively, the complex-\ac{NN} module should learn beam patterns that can effectively capture the characteristics of the underlying environment. The DeepMIMO O1\_28 and O1\_28B environments provide a good case study. Both environments feature similar topologies where a roadside \ac{BS} serves \acp{UE} located on the street. While all \acp{UE} are \ac{LOS} in the O1\_28 environment, a significant portion of the \acp{UE} are \ac{NLOS} due to blockage by a metal screen placed in front of the \ac{BS} in the O1\_28B environment. In order to provide coverage to the \ac{NLOS} \acp{UE}, the \ac{BS} in the O1\_28B environment needs to steer beams towards the two reflectors on both sides of the street. The majority of the \ac{LOS} \ac{UE} are also distributed on both sides of the \ac{BS}. 

\begin{figure*}%
\centering
\subfloat[O1\_28, $N_{\mathbf{W}}=6$]{\includegraphics[width=0.25\columnwidth,trim=4.5cm 1.0cm 4.5cm 1.1cm,clip]{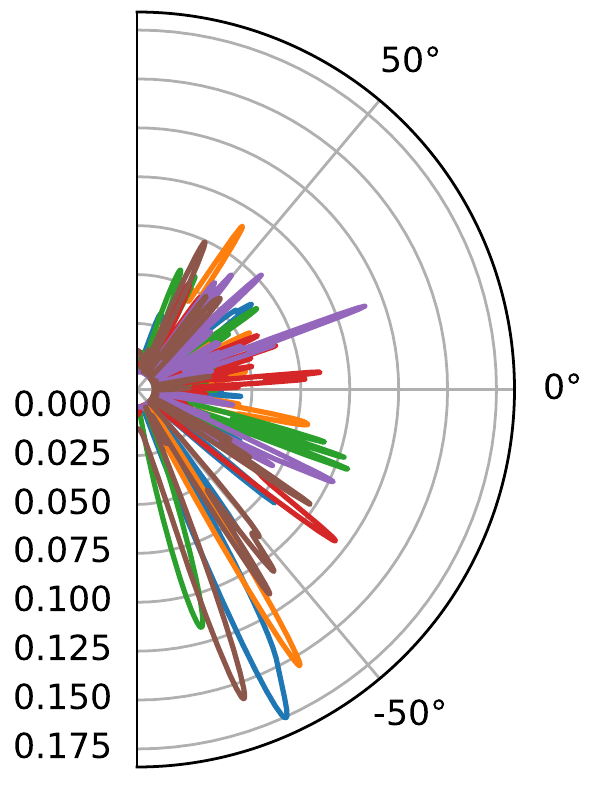}\label{figure:6_beam_codebook_o28}}
\subfloat[O1\_28, $N_{\mathbf{W}}=8$]{\includegraphics[width=0.25\columnwidth,trim=4.5cm 1.0cm 4.5cm 1.1cm,clip]{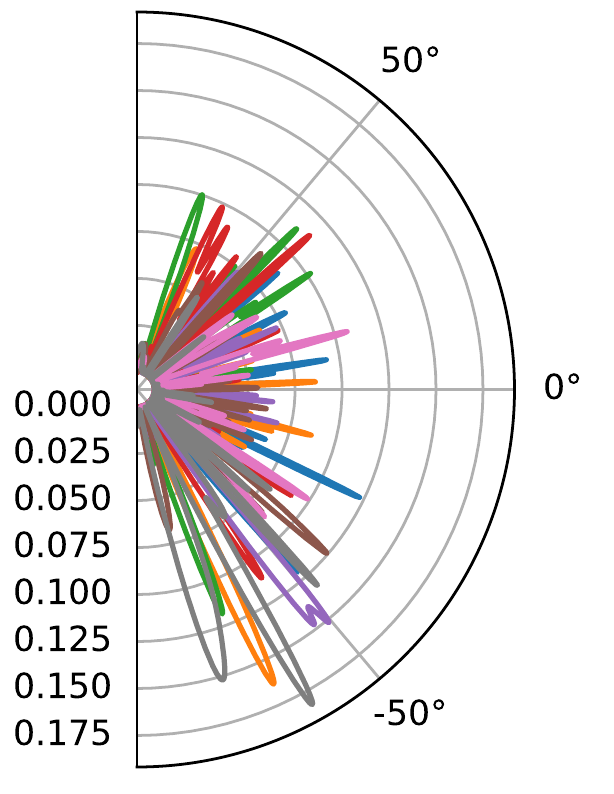}\label{figure:8_beam_codebook_o28}}
\subfloat[O1\_28, $N_{\mathbf{W}}=10$]{\includegraphics[width=0.25\columnwidth,trim=4.5cm 1.0cm 4.5cm 1.1cm,clip]{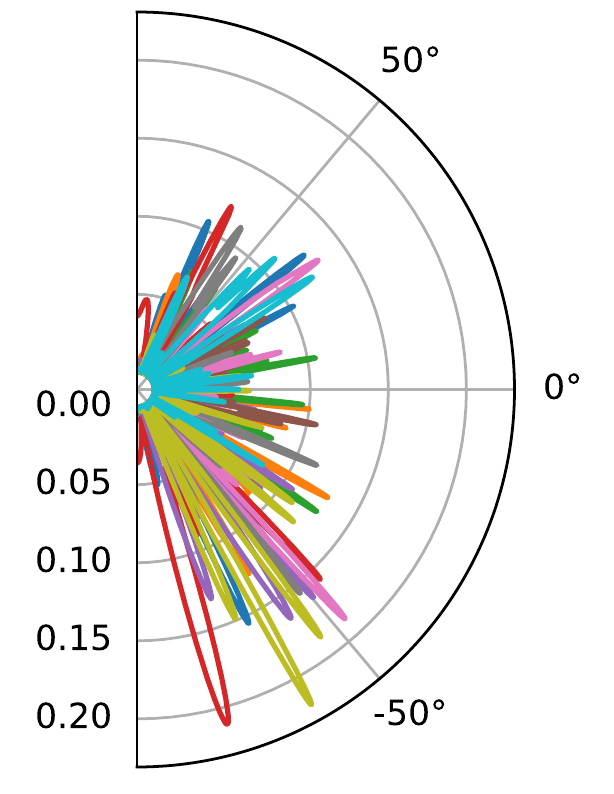}\label{figure:10_beam_codebook_o28}}
\subfloat[O1\_28, $N_{\mathbf{W}}=12$]{\includegraphics[width=0.25\columnwidth,trim=4.5cm 1.0cm 4.5cm 1.1cm,clip]{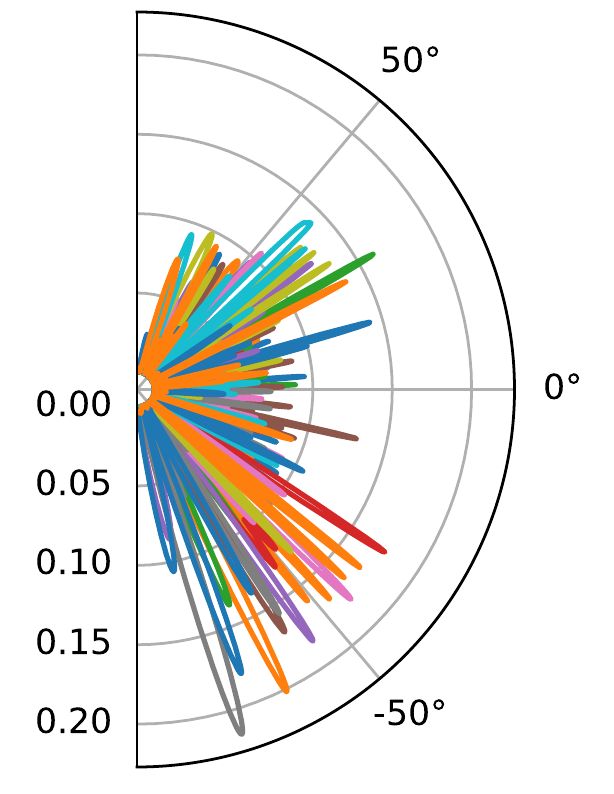}\label{figure:12_beam_codebook_o28}}
\hfill
\subfloat[O1\_28B, $N_{\mathbf{W}}=6$]{\includegraphics[width=0.25\columnwidth,trim=4.5cm 1.0cm 4.5cm 1.1cm,clip]{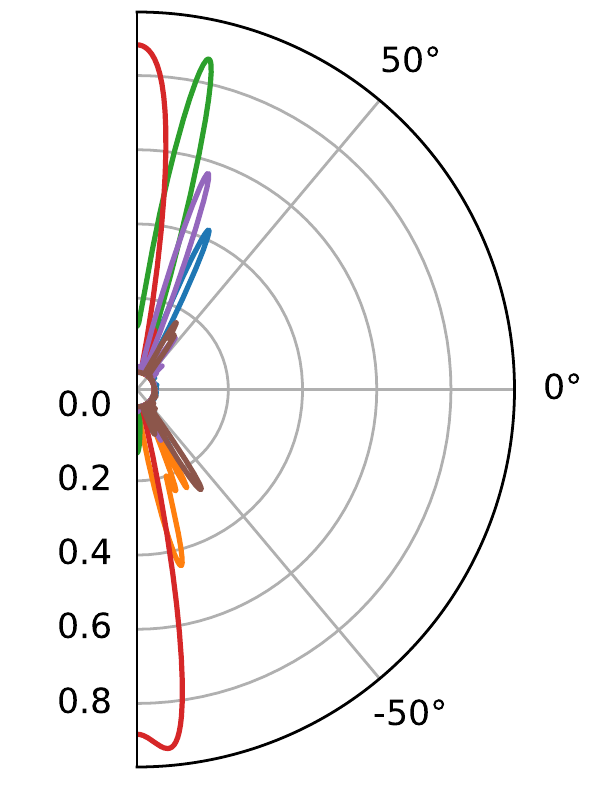}\label{figure:6_beam_codebook_o28b}}
\subfloat[O1\_28B, $N_{\mathbf{W}}=8$]{\includegraphics[width=0.25\columnwidth,trim=4.5cm 1.0cm 4.5cm 1.1cm,clip]{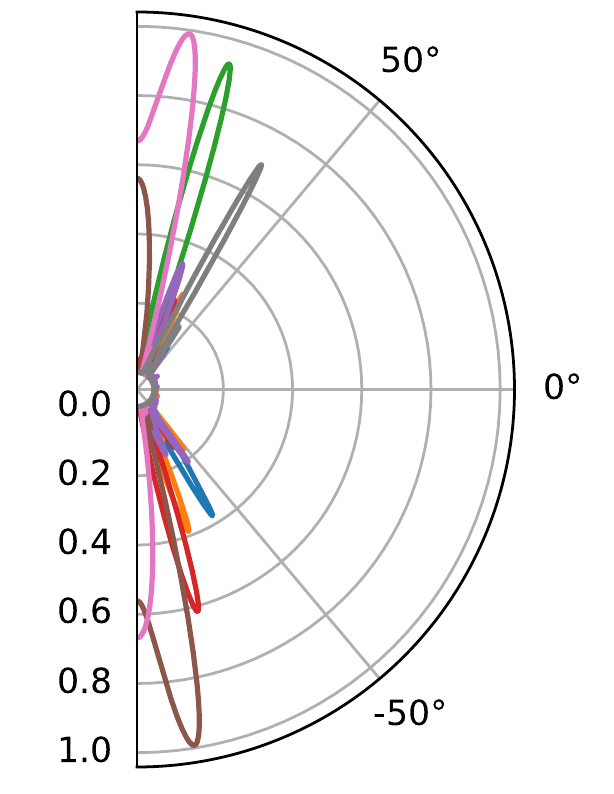}\label{figure:8_beam_codebook_o28b}}
\subfloat[O1\_28B, $N_{\mathbf{W}}=10$]{\includegraphics[width=0.25\columnwidth,trim=4.5cm 1.0cm 4.5cm 1.1cm,clip]{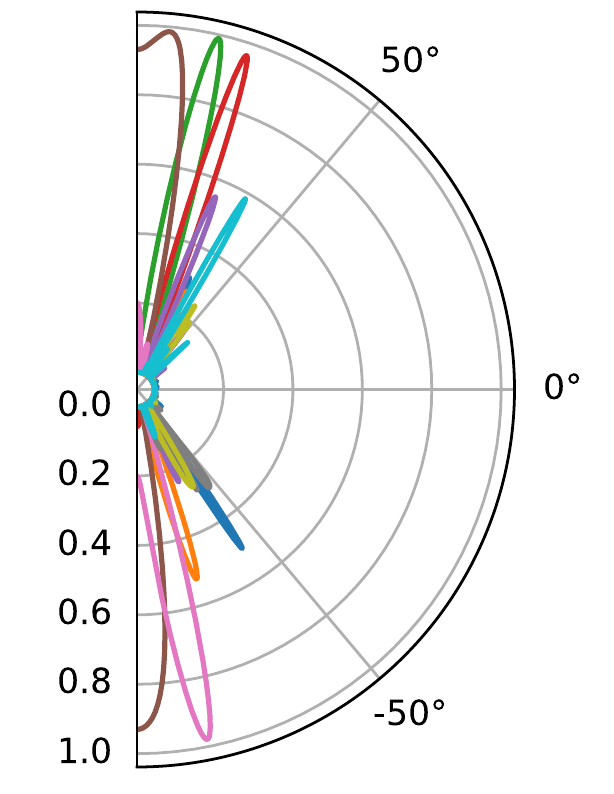}\label{figure:10_beam_codebook_o28b}}
\subfloat[O1\_28B, $N_{\mathbf{W}}=12$]{\includegraphics[width=0.25\columnwidth,trim=4.5cm 1.0cm 4.5cm 1.1cm,clip]{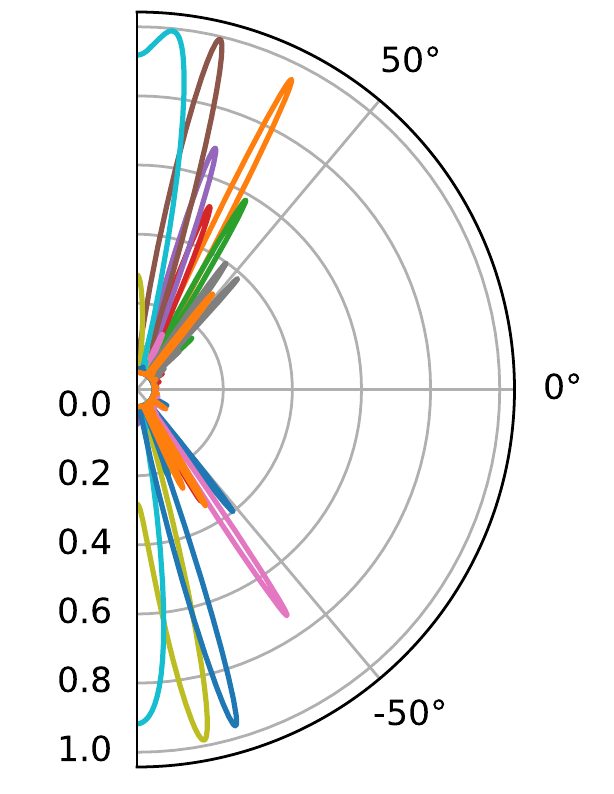}\label{figure:12_beam_codebook_o28b}}
\hfill
\caption{Learned beam patterns with different probing codebook sizes in the DeepMIMO O1\_28 and O1\_28B environments.}\label{figure:codebook_pattern}
\end{figure*}

The learned probing codebook patterns in both environments is shown in Fig. \ref{figure:codebook_pattern}. Firstly, the learned radiation patterns are similar with increasing probing codebook sizes $N_{\mathbf{W}}$ in either environment. Regardless of the codebook size, the probing codebook consistently learns to focus energy on specific areas. This indicates that the complex-\ac{NN} module can consistently learn probing codebook patterns in a given environment. 
Unlike conventional \ac{DFT} beams which have a single main lobe, the learned beams often have multiple main lobes. Such beam patterns can likely capture more information about the propagation environment given the small number of probing beams allowed. 
Furthermore, the learned probing codebooks are adapted to the particular characteristics of different environments. The learned beam patterns in the O1\_28 environment are drastically different from those in O1\_28B. In the O1\_28B environment, the codebooks are optimized to focus energy in the angular regions close to $\pm 90^{\circ}$, corresponding to the positions of the \ac{LOS} \acp{UE} and the reflectors. In comparison, the codebooks learned in the O1\_28 environment spread the energy much more evenly in the broadside direction, which is consistent with the even distribution of \ac{LOS} \acp{UE} in front of the \ac{BS} in this environment.
While the complex-\ac{NN} module is not explicitly optimized to leverage spatial patterns of the environment, it can nevertheless consistently learn probing beams that captures particular characteristics of the environment.

\subsection{Does the proposed method achieve lower beam sweeping complexity?}

With the proposed beam alignment method, all \acp{UE} can measure the probing beams simultaneously when the \ac{BS} sweeps the probing codebook. If the \ac{BS} choose to sweep the top-$k$ predicted beams, those beams may be different for each UE. Hence the beam sweeping complexity is $N_{\mathbf{W}}+K \cdot k_{\mathbbm{1}_{\{k>1\}}}$ for $K$ \acp{UE}. Each \ac{UE} needs to feedback the received signal power of the $N_\mathbf{W}$ probing beams. If the \ac{BS} choose to sweep additional beams, each \ac{UE} only needs to feedback the index of the best beam. With the 2-tier hierarchical beam search, the 1st-tier wide beams can be transmitted using \acp{SSB} and be measured by all \acp{UE} simultaneously, while different 2nd-tier children beams need to be swept for each \ac{UE}. On average, the beam sweeping complexity is $N_{\mathbf{W}} + K \frac{N_{\mathbf{V}}}{N_{\mathbf{W}}}$ for $K$ \acp{UE}. Each \ac{UE} needs to feedback the index of the best beam in each tier. With the binary hierarchical beam search, the two first layer beams can be measured simultaneously by all \acp{UE} while the subsequent beam sweeping needs to be done for each different \ac{UE}. Hence the beam sweeping complexity is $2+2K\log_{2}\frac{N_{\mathbf{V}}}{2}$ for $K$ \acp{UE}. Each \ac{UE} needs to feedback the index of the best beam in each level of the binary search. With the exhaustive beam search, the $N_{\mathbf{V}}$ beams can be measured by all UEs simultaneously. The beam complexity is $N_{\mathbf{V}}$ regardless of the number of \acp{UE}. Each \ac{UE} needs to feedback the index of the best beam. 
A summary of the beam sweeping and feedback complexity of the proposed method and the baselines is shown in Table \ref{table:BScomplexity}.

\begin{table*}
\centering
  \caption{Beam Sweeping Complexity for $K$ UEs}\label{table:BScomplexity}
    \begin{tabular}{| c | c | c |}
    \hline
    \textbf{Beam alignment method} & \textbf{Beam sweeping complexity} & \textbf{Feedback complexity} \\ \hline
    Proposed method & $N_{\mathbf{W}} + K \cdot k_{\mathbbm{1}_{\{k>1\}}}$ & \makecell{$K N_{\mathbf{W}}$ received signal power\\ + $K \cdot \mathbbm{1}_{\{k>1\}}$ beam indices} \\ \hline
    2-tier hierarchical search & $N_{\mathbf{W}} + K \frac{N_{\mathbf{V}}}{N_{\mathbf{W}}}$ & $2 K$ beam indices \\ \hline
    Binary hierarchical search & $2+2 K \log_{2}\frac{N_{\mathbf{V}}}{2}$ & $K \log_{2}N_{\mathbf{V}}$ beam indices \\ \hline
    Exhaustive search & $N_{\mathbf{V}}$ & $K$ beam indices\\ \hline
    \end{tabular}
\end{table*}

\begin{figure*}%
\centering
\subfloat[1 UE]{\includegraphics[width=0.5\columnwidth,trim=0.3cm 0.3cm 1.2cm 1.3cm,clip]{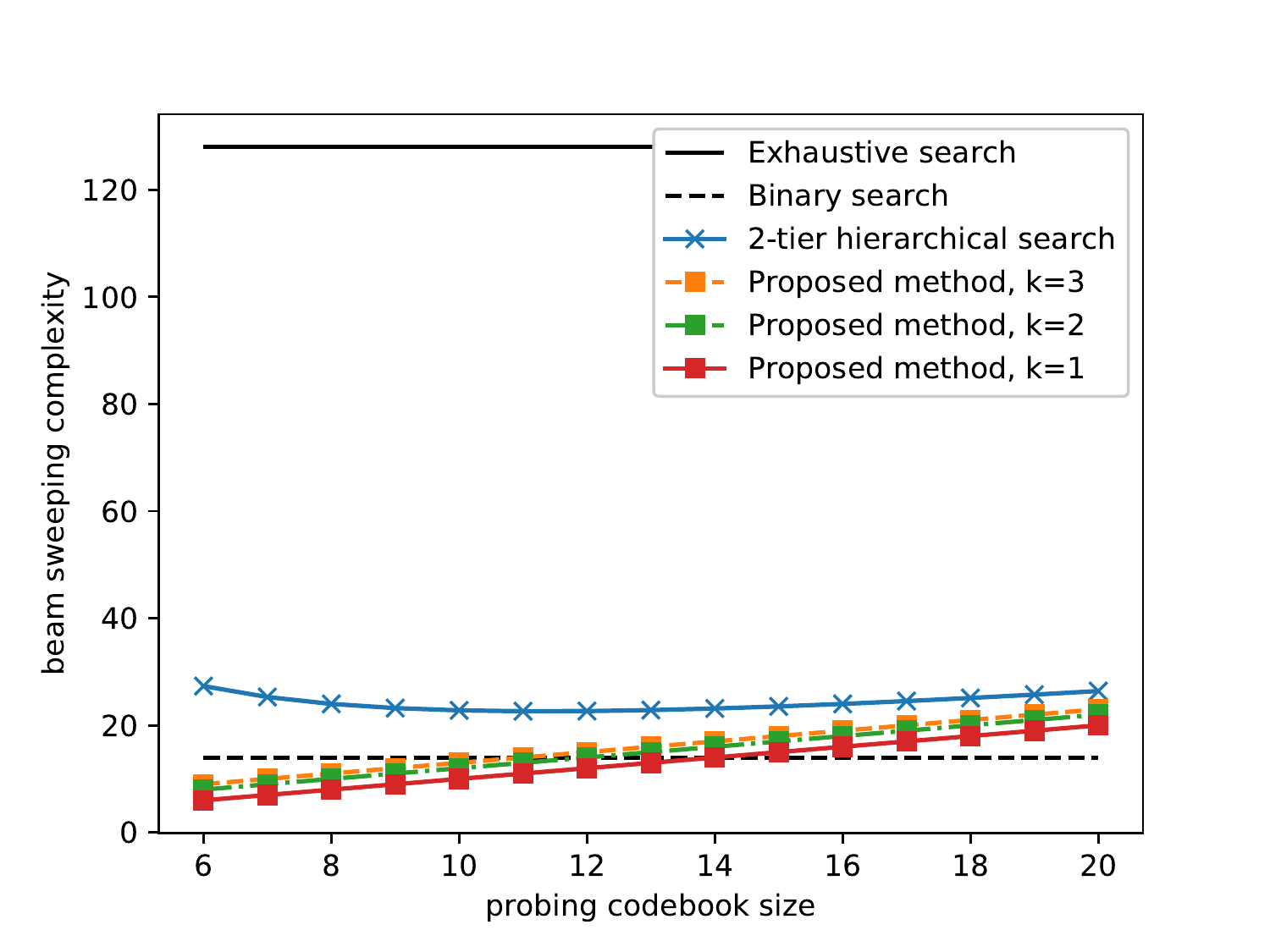}\label{figure:beam_sweeping_complexity_1ue}}
\subfloat[5 UEs]{\includegraphics[width=0.5\columnwidth,trim=0.3cm 0.3cm 1.2cm 1.3cm,clip]{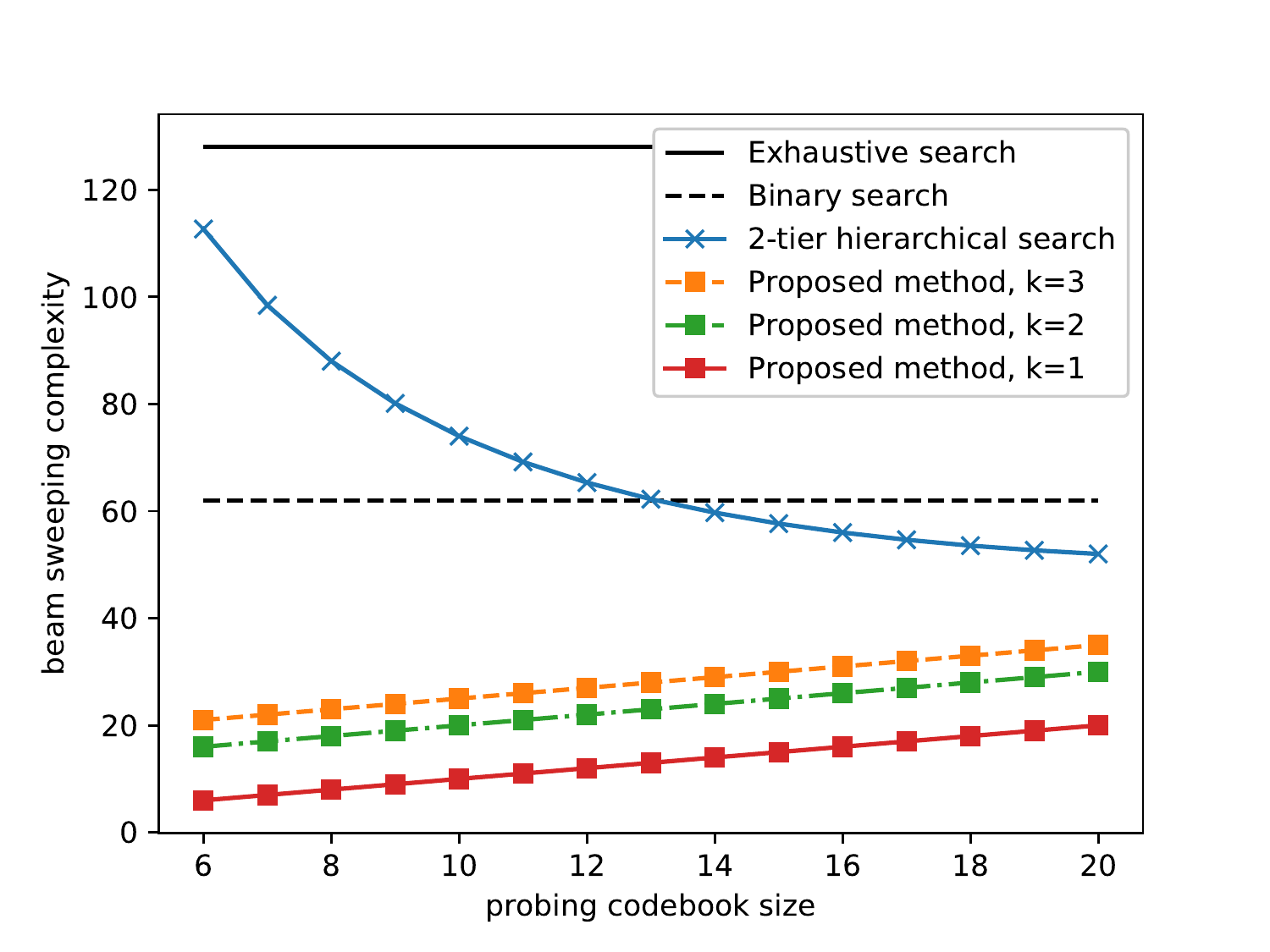}\label{figure:beam_sweeping_complexity_5ue}}
\hfill
\subfloat[10 UEs]{\includegraphics[width=0.5\columnwidth,trim=0.3cm 0.3cm 1.2cm 1.3cm,clip]{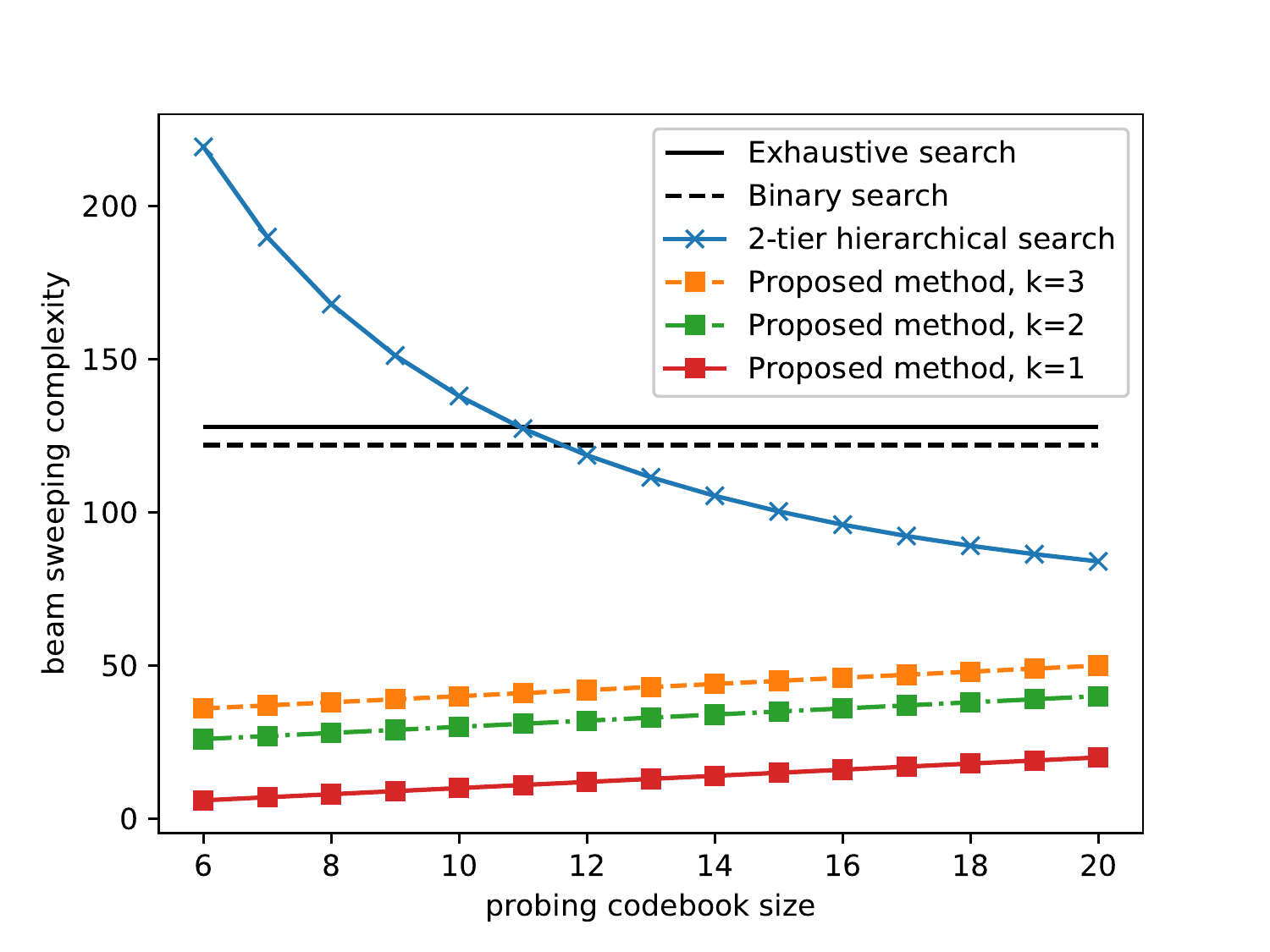}\label{figure:beam_sweeping_complexity_10ue}}
\subfloat[15 UEs]{\includegraphics[width=0.5\columnwidth,trim=0.3cm 0.3cm 1.2cm 1.3cm,clip]{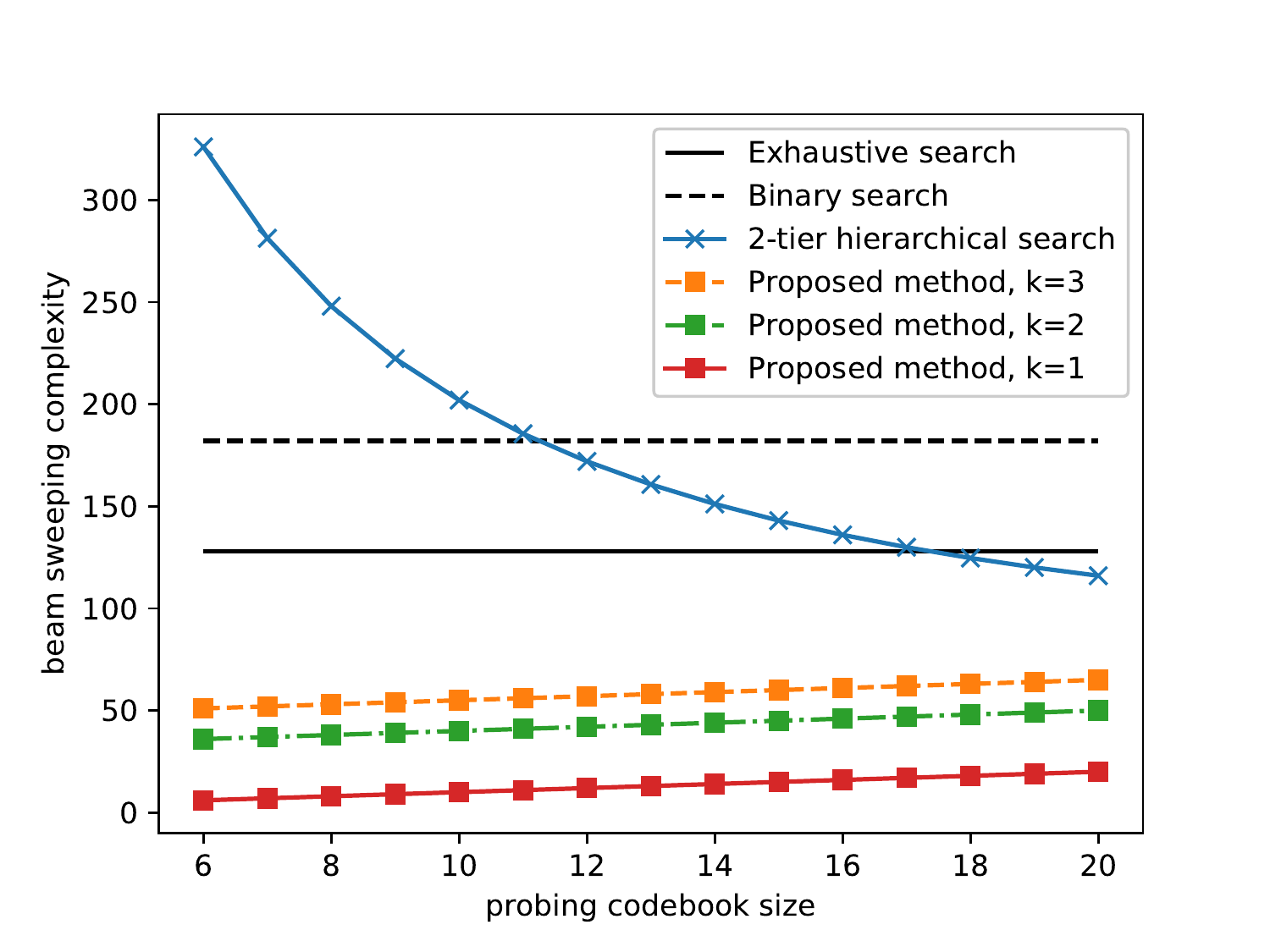}\label{figure:beam_sweeping_complexity_15ue}}
\caption{Beam sweeping complexity vs. probing codebook size.}\label{figure:beam_sweeping_complexity}
\end{figure*}

A comparison of the beam sweeping complexity with 1,5,10 and 15 \acp{UE} is shown in Fig.\ref{figure:beam_sweeping_complexity}. When considering a single \ac{UE}, the proposed method has lower beam sweeping complexity compared to the exhaustive search and the 2-tier hierarchical beam search. With fewer than 11 probing beams, the proposed method also incurs lower beam sweeping complexity than the binary beam search does even when sweeping 2 or 3 additional beams. When considering simultaneous beam alignment for multiple \acp{UE} such as 5, 10 and 15 \acp{UE}, the beam sweeping complexity of the proposed method is lower than that of any baseline. 
In the 2 \ac{NLOS} environments (DeepMIMO I3 and O1\_28B), when considering simultaneous beam alignment for 10 \acp{UE}, the proposed method with 12 probing beams and $k=3$ can achieve an average SNR similar to that of the exhaustive beam search, at least 3.65 dB better than that of the 2-tier hierarchical beam search and at least 7.71 dB better than that of the binary beam search, while incurring less than 35.4\% of the beam sweeping complexity of any baseline. 
With 12 probing beams but without additional beam sweeping ($k=1$), the proposed method can still beat the hierarchical search baselines while reducing the beam sweeping complexity by 10$\times$. 

\subsection{Is the proposed beam alignment method robust to noise?}
\begin{figure*}%
\centering
\subfloat[Rosslyn]{\includegraphics[width=0.5\columnwidth,trim=0.3cm 0.3cm 1.2cm 1.3cm,clip]{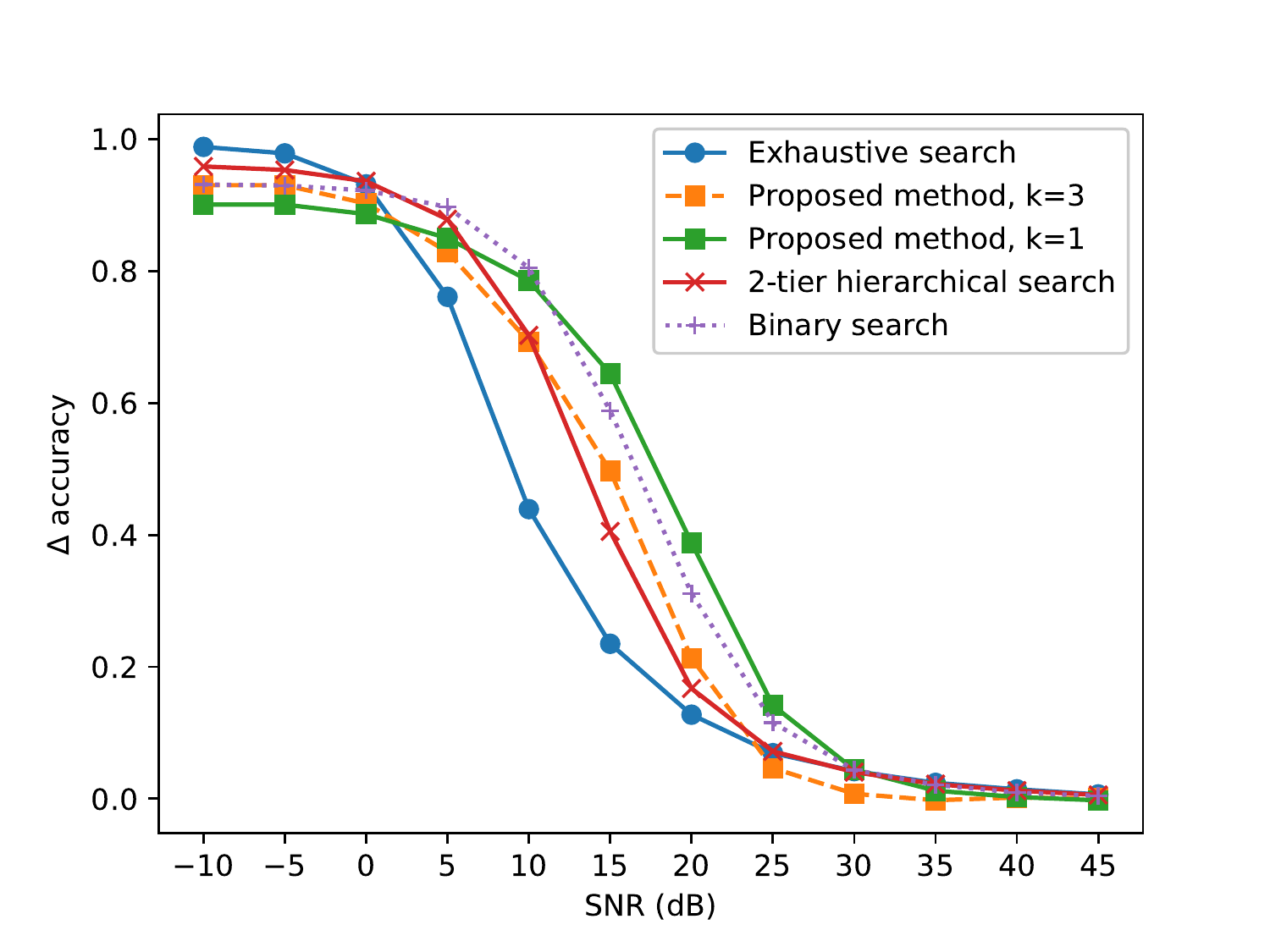}\label{figure:acc_vs_noise_rosslyn}}
\subfloat[DeepMIMO O1\_28]{\includegraphics[width=0.5\columnwidth,trim=0.3cm 0.3cm 1.2cm 1.3cm,clip]{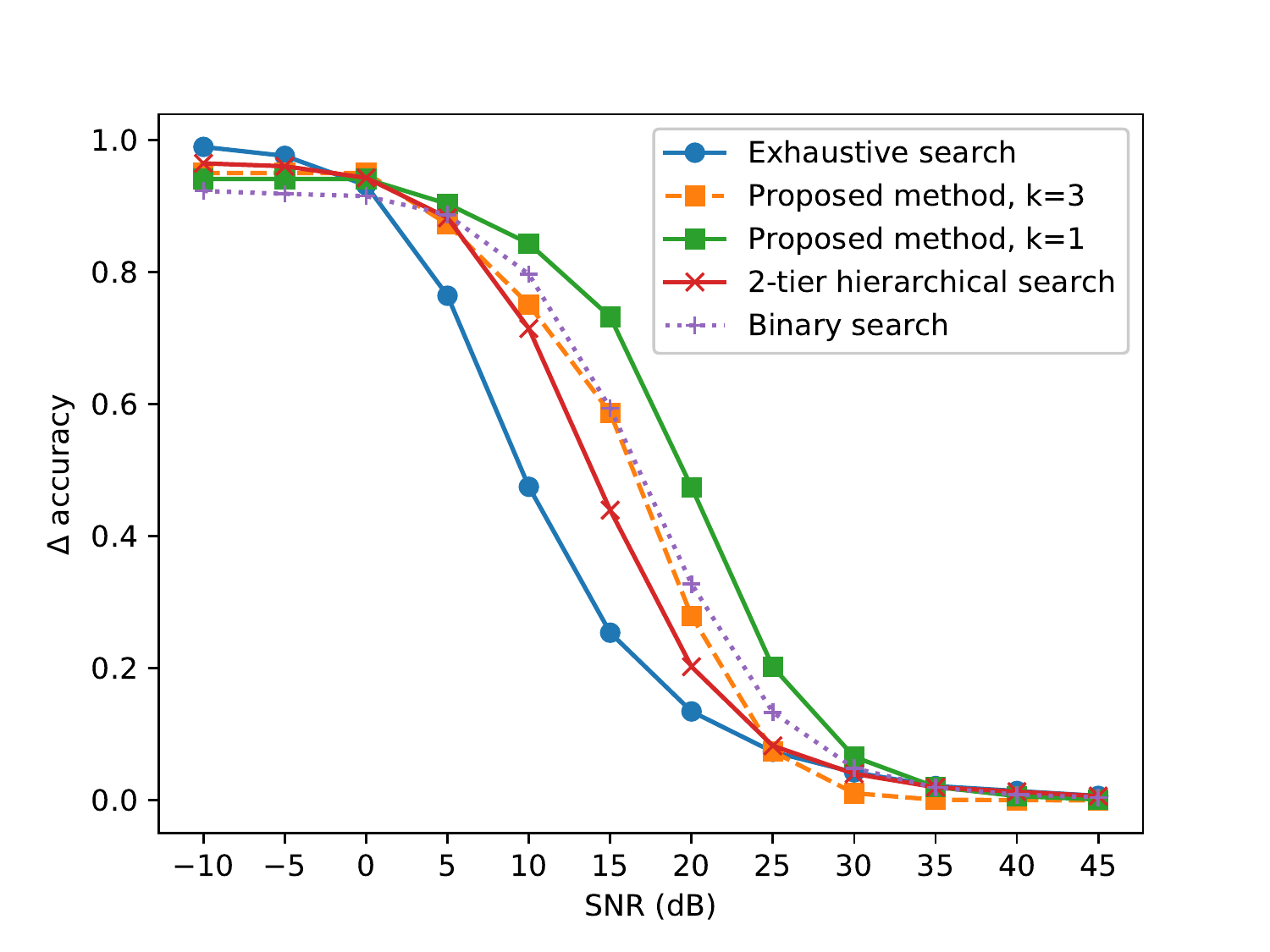}\label{figure:acc_vs_noise_O1_28}}
\hfill
\subfloat[DeepMIMO I3]{\includegraphics[width=0.5\columnwidth,trim=0.3cm 0.3cm 1.2cm 1.3cm,clip]{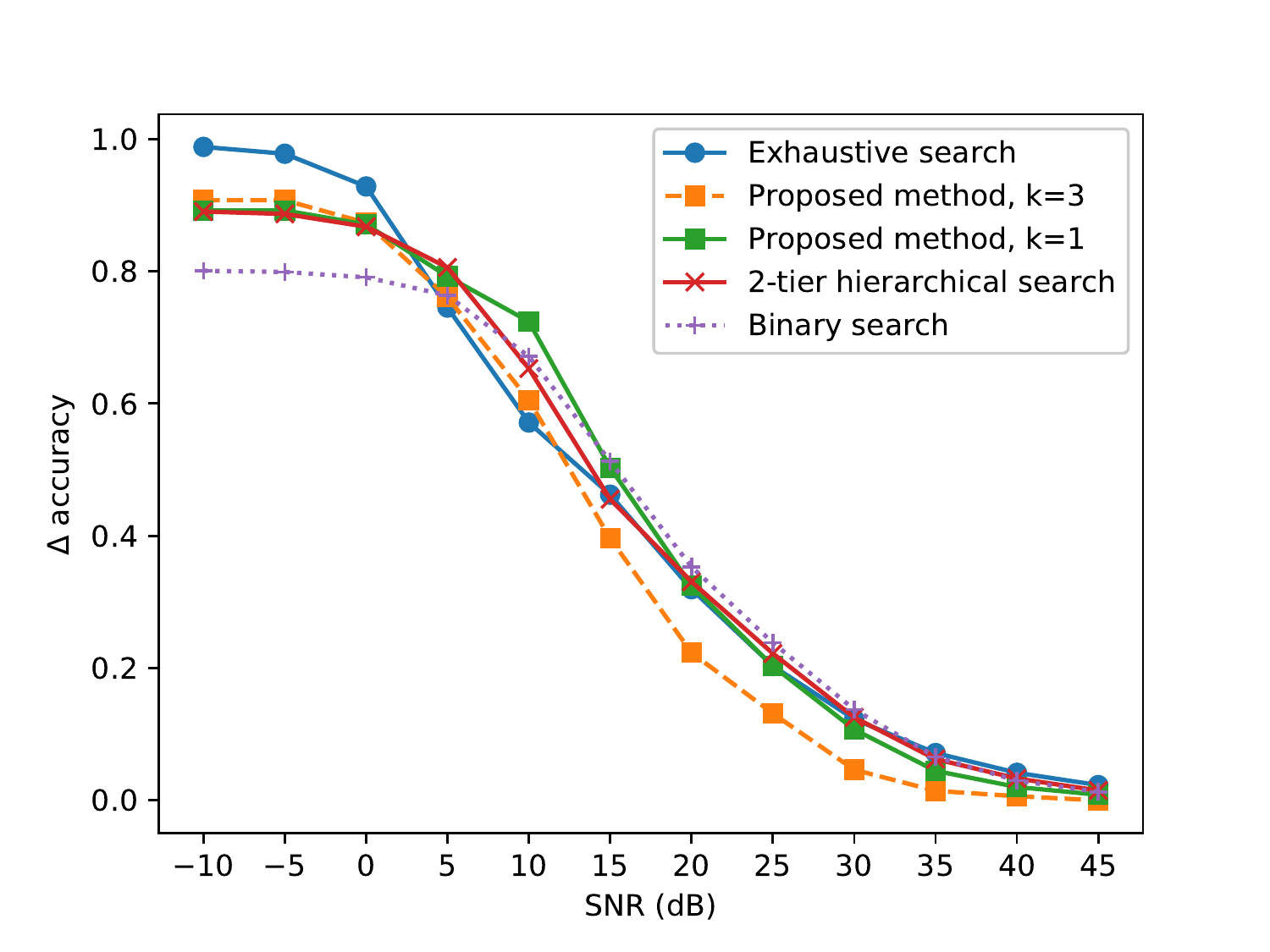}\label{figure:acc_vs_noise_I3}}
\subfloat[DeepMIMO O1\_28B]{\includegraphics[width=0.5\columnwidth,trim=0.3cm 0.3cm 1.2cm 1.3cm,clip]{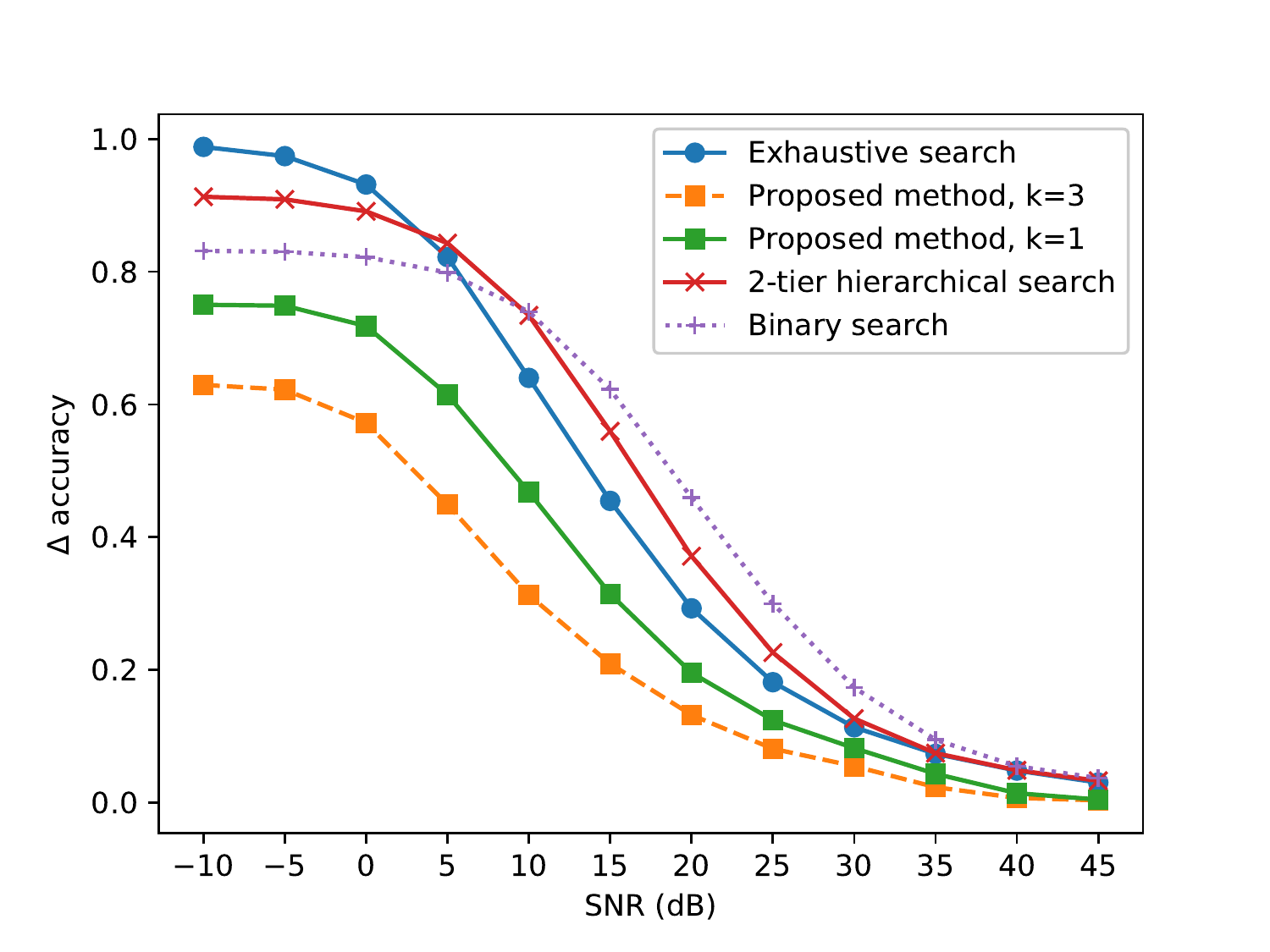}\label{figure:acc_vs_noise_O1_28B}}
\caption{Beam alignment accuracy gap vs. SNR. The vertical axis represents the accuracy degradation from when there is no noise in the received signal. The horizontal axis represents the average SNR of the optimal narrow beams of all UEs.}\label{figure:acc_vs_snr}
\end{figure*}

The proposed beam alignment method, like any beam sweeping-based approach, relies on measurements of the received power of the \ac{BF} signals. As a result, noise in the received \ac{BF} signal may have significant impacts on the beam alignment performance. We compare the beam alignment accuracy at various noise levels to that when there is no noise. 
The accuracy degradation is defined as the absolute difference between the accuracy with no noise and the accuracy at a certain \ac{SNR} level. The accuracy degradation at various \ac{SNR} levels is shown in Fig. \ref{figure:acc_vs_snr}. 
In the \ac{LOS} environments (Rosslyn and DeepMIMO O1\_28), the accuracy degradation of all compared methods is minimal when the \ac{SNR} is over 35 dB. When the \ac{SNR} is between 5 dB and 25 dB, the exhaustive search experiences the least amount of accuracy drop. The proposed method with $k$ = 3 experiences similar levels of degradation compared to the hierarchical search baselines. 
In the \ac{NLOS} environments (DeepMIMO I3 and O1\_28B), the accuracy degradation is much more noticeable even at high \ac{SNR} levels of over 35 dB. The proposed method also performs more favorably in terms of accuracy drop. With $k$ = 3, it experiences the least amount of accuracy degradation when the \ac{SNR} is over 15 dB in the DeepMIMO I3 environment. In the O1\_28B environment, the proposed method experiences less degradation than any other baseline at all \ac{SNR} levels. 

\subsection{How does the learned probing codebook help the beam predictor?}\label{section:eval_explanation}
\begin{figure*}%
\centering
\subfloat[Rosslyn]{\includegraphics[width=0.5\columnwidth,trim=0.3cm 0.3cm 1.2cm 1.3cm,clip]{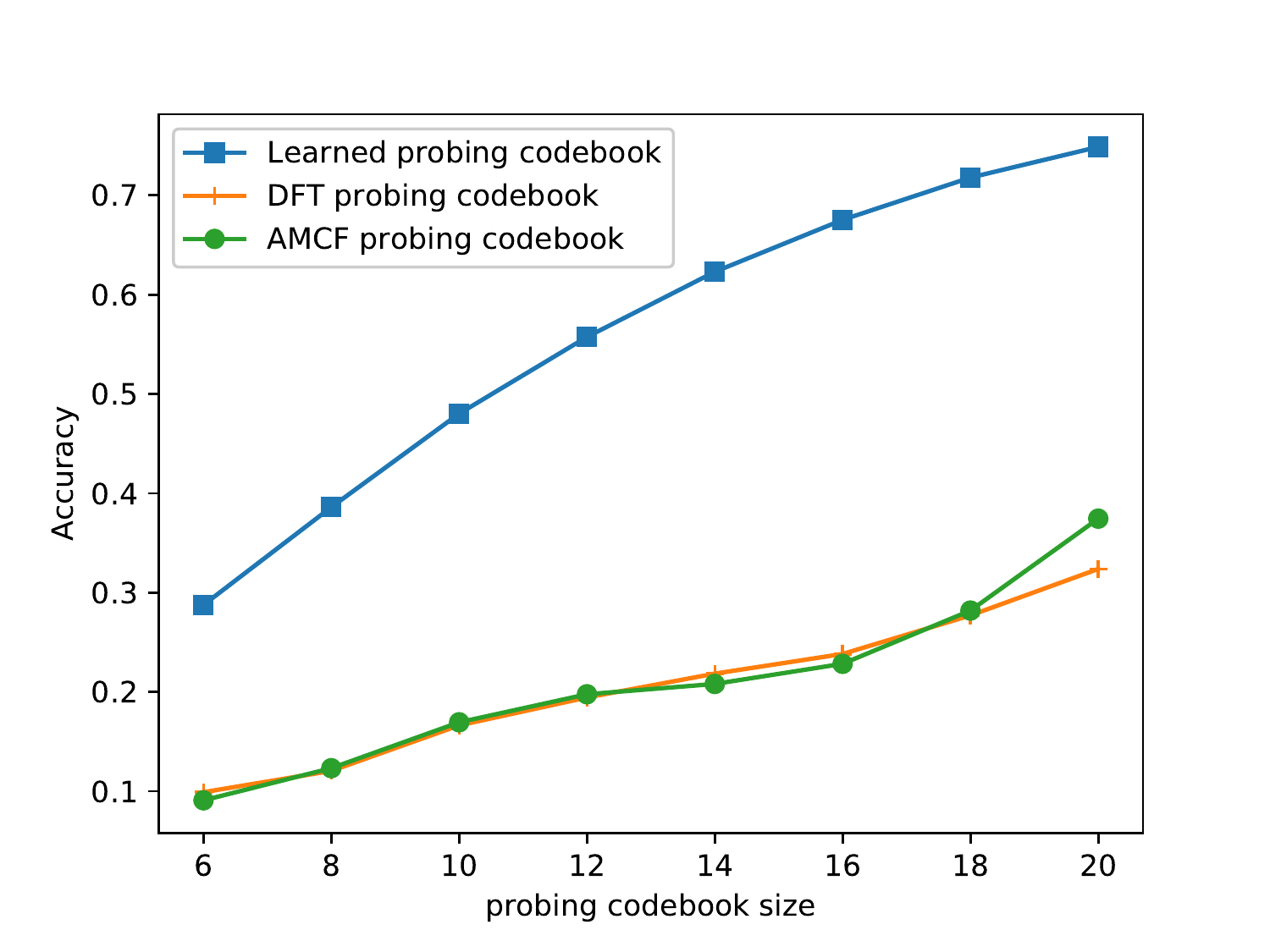}\label{figure:acc_vs_codebook_rosslyn}}
\subfloat[DeepMIMO O1\_28]{\includegraphics[width=0.5\columnwidth,trim=0.3cm 0.3cm 1.2cm 1.3cm,clip]{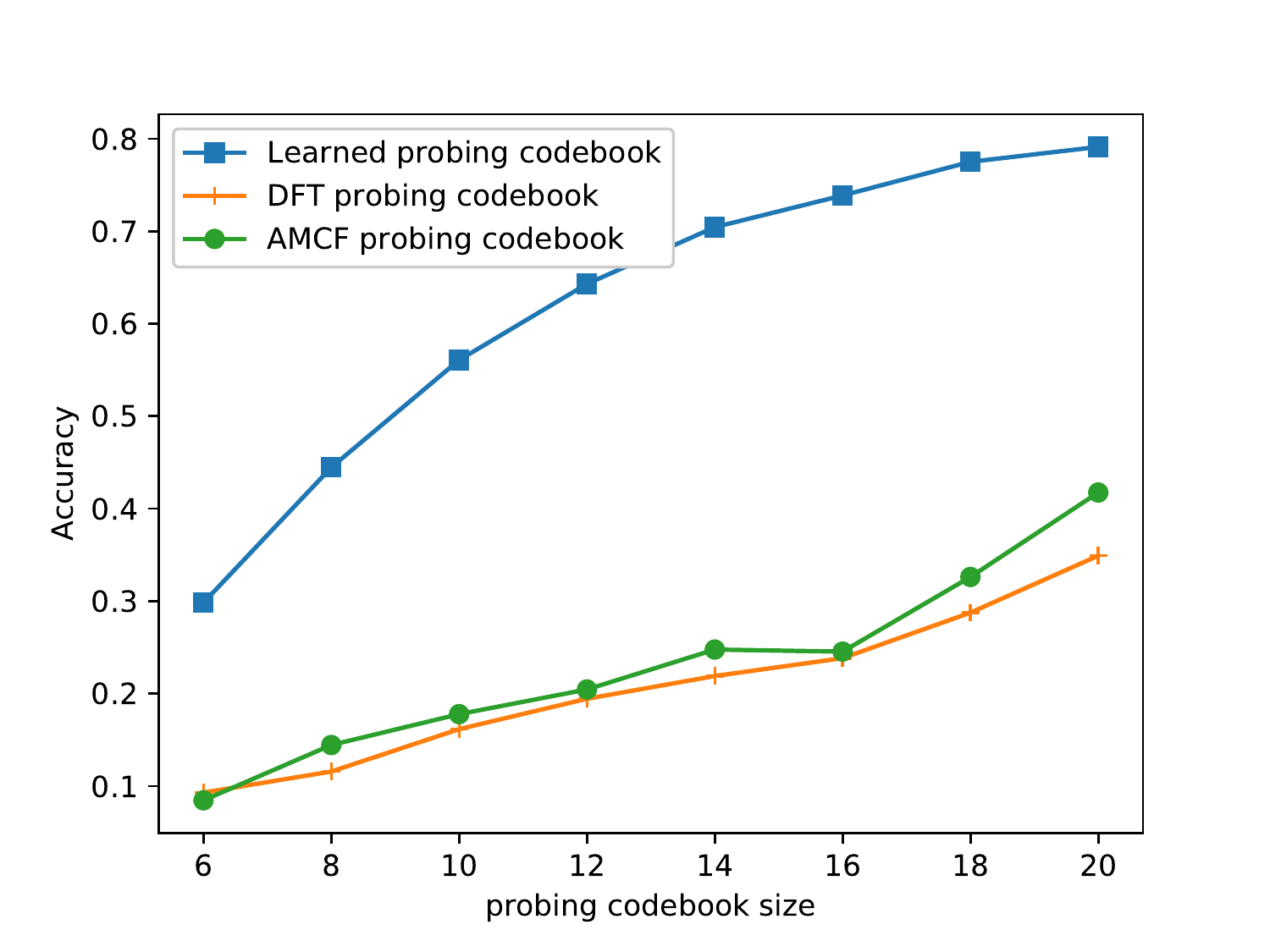}\label{figure:acc_vs_codebook_O1_28}}
\hfill
\subfloat[DeepMIMO I3]{\includegraphics[width=0.5\columnwidth,trim=0.3cm 0.3cm 1.2cm 1.3cm,clip]{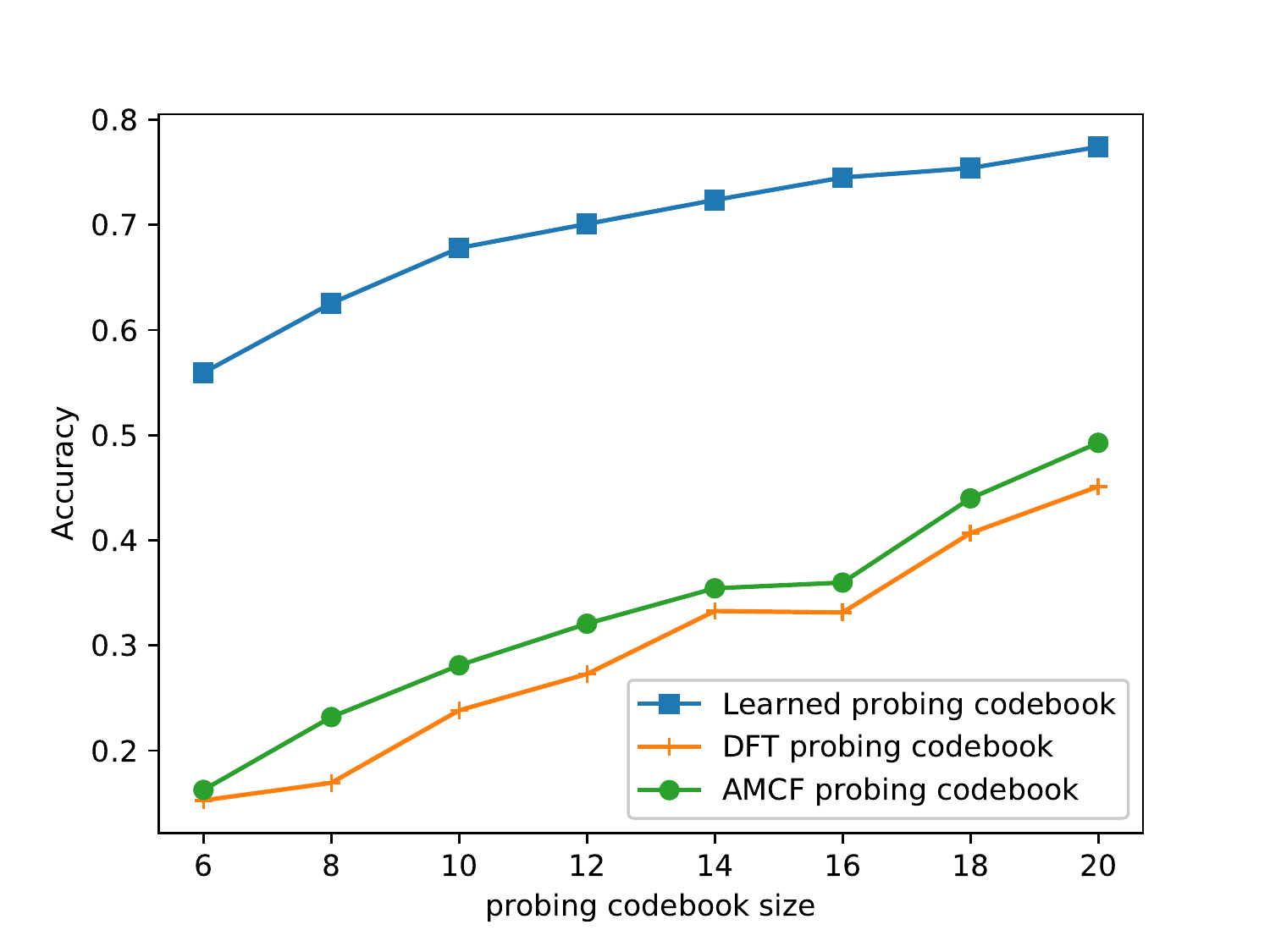}\label{figure:acc_vs_codebook_I3}}
\subfloat[DeepMIMO O1\_28B]{\includegraphics[width=0.5\columnwidth,trim=0.3cm 0.3cm 1.2cm 1.3cm,clip]{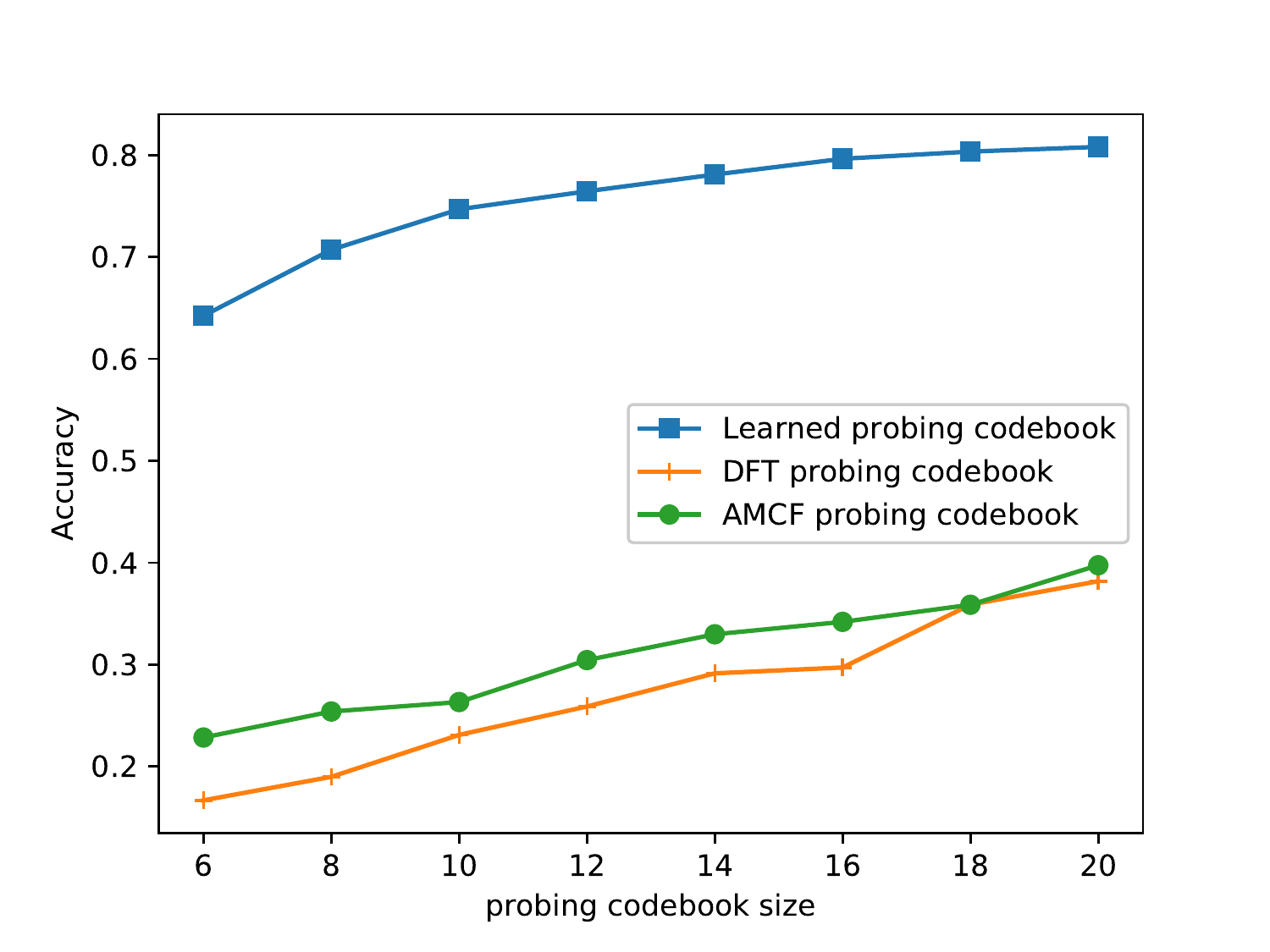}\label{figure:acc_vs_codebook_O1_28B}}
\caption{Accuracy vs. probing codebook size.}\label{figure:acc_vs_codebook}
\end{figure*}

The trainable probing codebook is an important part of the proposed bean alignment method and should be optimized to help the subsequent optimal-beam classification task. To verify this, the performance of the \ac{MLP} classifier is evaluated while the trainable probing codebook is replaced with a predetermined probing codebook. Two predetermined probing codebooks are considered: a \ac{DFT} codebook with $N_{\mathbf{W}}$ evenly-spaced narrow beams which is similar to the sparse codebook used in \cite{ma2020MLbeamalignment}, and a wide-beam codebook whose $N_{\mathbf{W}}$ evenly-spaced wide beams are generated using the \ac{AMCF} algorithm. The \ac{MLP} is trained from scratch using the received signal power of each predetermined probing codebook. A beam alignment accuracy comparison of the learned probing codebook and the predetermined ones in all 4 environments is shown in Fig. \ref{figure:acc_vs_codebook}. 
In all 4 environments, the learned probing codebook achieves significantly better beam alignment accuracy. By placing beams strategically according to the propagation environment instead of evenly in the angular space regardless of the environment, the learned probing codebook is much more effective at capturing channel characteristics, which greatly benefits the downstream classification task. 

The probing codebook can also be viewed through the lens of data clustering and representation learning, which provides a further explanation of how the learned codebook help select the optimal beam. By performing beam sweeping using the probing codebook, the high-dimensional channel vector $\mathbf{h} \in \mathbbm{C}^{N_{t} \times 1}$ is transformed into a feature vector of received signal power values $\mathbf{x} \in \mathbbm{R}^{N_{\mathbf{W}} \times 1}$ lying in a lower-dimensional subspace determined by the probing codebook. If the transformed feature vectors with the same optimal narrow beam are assigned to the same cluster, a good probing codebook should intuitively make clusters corresponding to different narrow beams well separated so that the \ac{MLP} can more easily predict the optimal beam. After beam sweeping using the probing codebook, channel realizations with the same optimal narrow beam should be close to each other in the transformed subspace, while those with different optimal narrow beams should be farther apart. This is similar to the representation learning problem in \ac{ML}, which often seeks to learn low-dimensional representations of high-dimensional data that exhibits natural clustering according to the data labels \cite{bengio2013representation}.
One measure of the clustering quality is the silhouette coefficient \cite{rousseeuw1987silhouettes}. For a dataset $\mathcal{D}$, its silhouette coefficient $S(\mathcal{D}) \in [-1,1]$ is
\begin{equation}
    S(\mathcal{D}) = \mathop{\mathbb{E}}_{i\in\mathcal{D}}\Big[\frac{b(i)-a(i)}{\max\{a(i),b(i)\}}\Big],
\end{equation}
where $a(i)$ is the mean intra-cluster distance of a data sample $i$ and $b(i)$ is its distance to the nearest cluster of which it is not a part of. A higher silhouette coefficient indicates better clustering and better separability of the data, which will likely make classifying the optimal beam easier. The silhouette coefficients of the received signal power vector $\mathbf{x}$ in all 4 environments are shown in Table \ref{table:silhouette_coefficients}. Compared to the predefined \ac{AMCF} and \ac{DFT} probing codebooks, the learned codebook consistently achieves better silhouette coefficients regardless of the environment and the number of probing beams, thus explaining its superior beam alignment performance.

\begin{figure*}%
\centering
\subfloat[Rosslyn, learned]{\includegraphics[width=0.33\columnwidth,trim=0.2cm 0.2cm 0.2cm 0.2cm,clip]{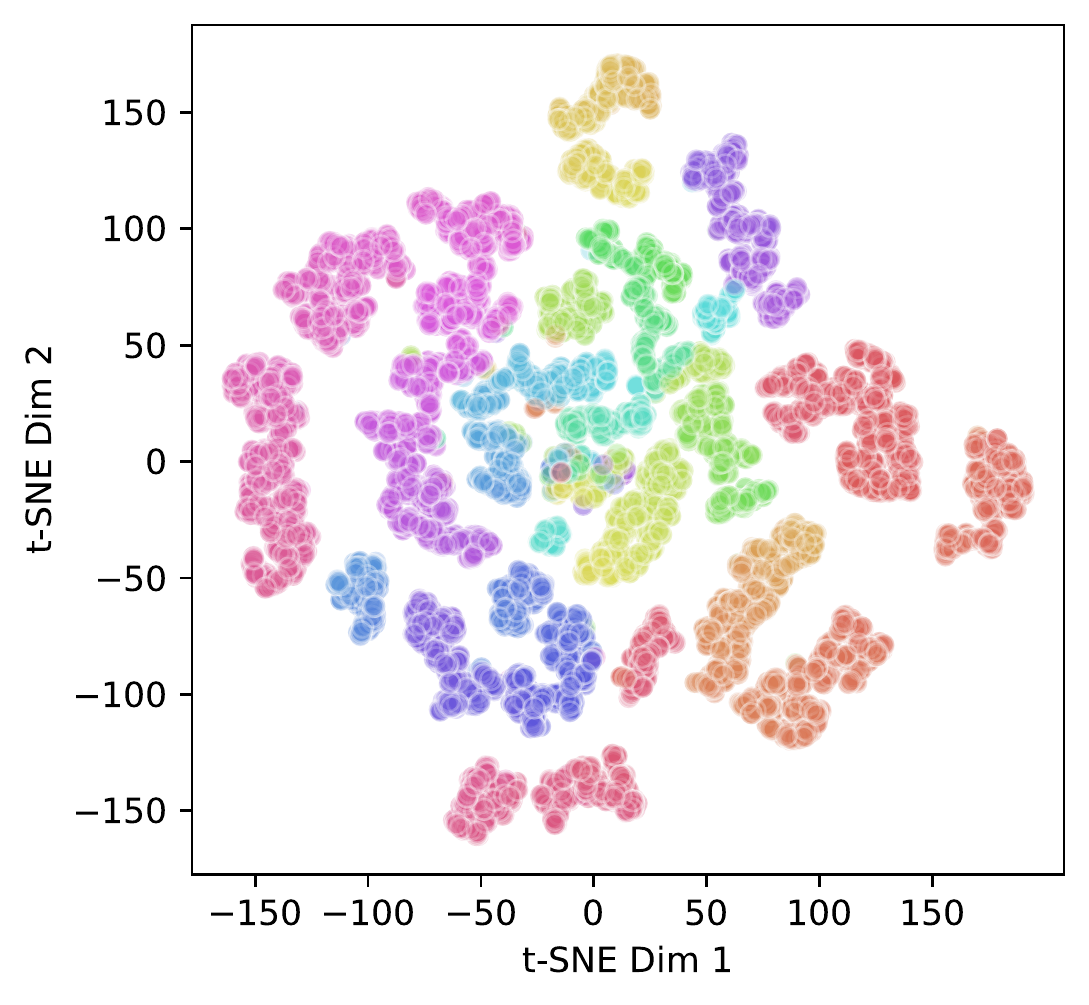}\label{figure:tsne_learned_rosslyn}}
\subfloat[Rosslyn, AMCF]{\includegraphics[width=0.33\columnwidth,trim=0.2cm 0.2cm 0.2cm 0.2cm,clip]{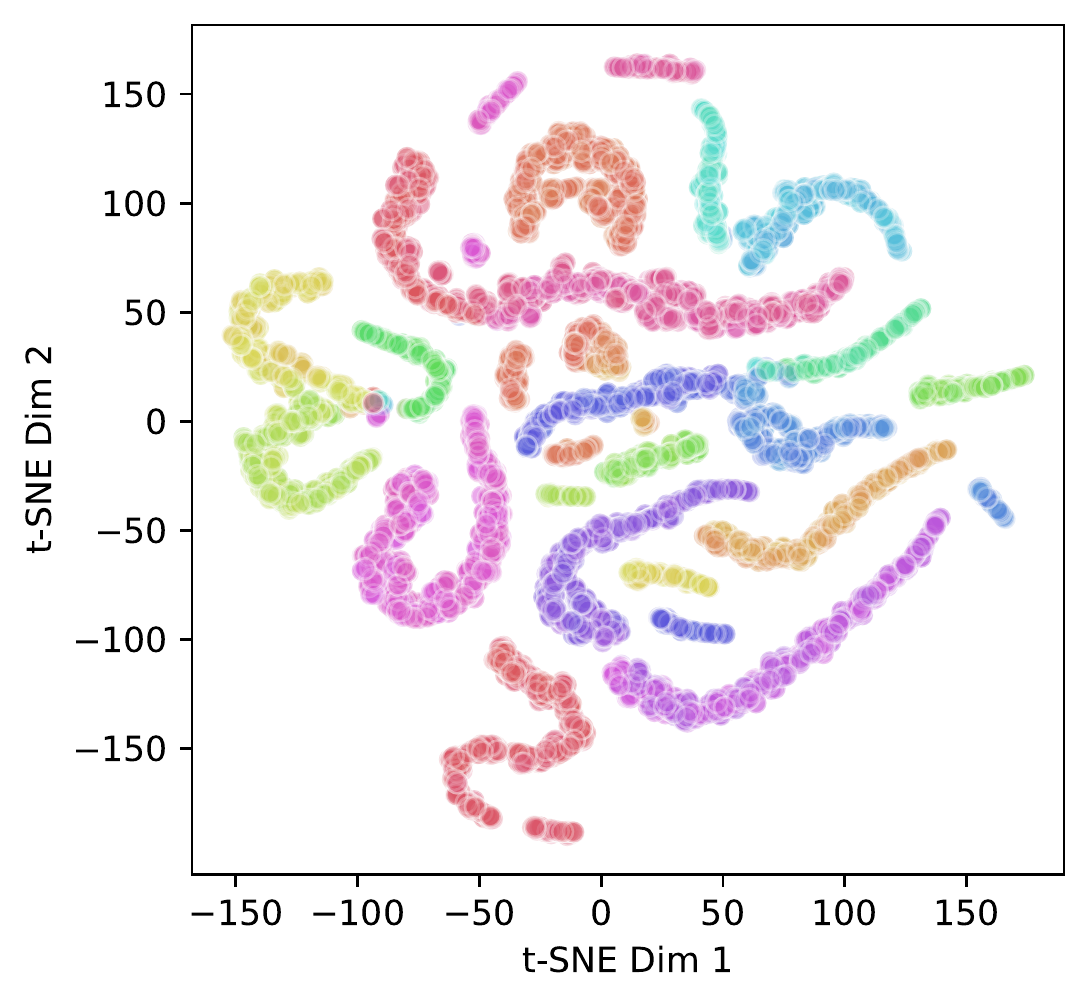}\label{figure:tsne_AMCF_rosslyn}}
\subfloat[Rosslyn, DFT]{\includegraphics[width=0.33\columnwidth,trim=0.2cm 0.2cm 0.2cm 0.2cm,clip]{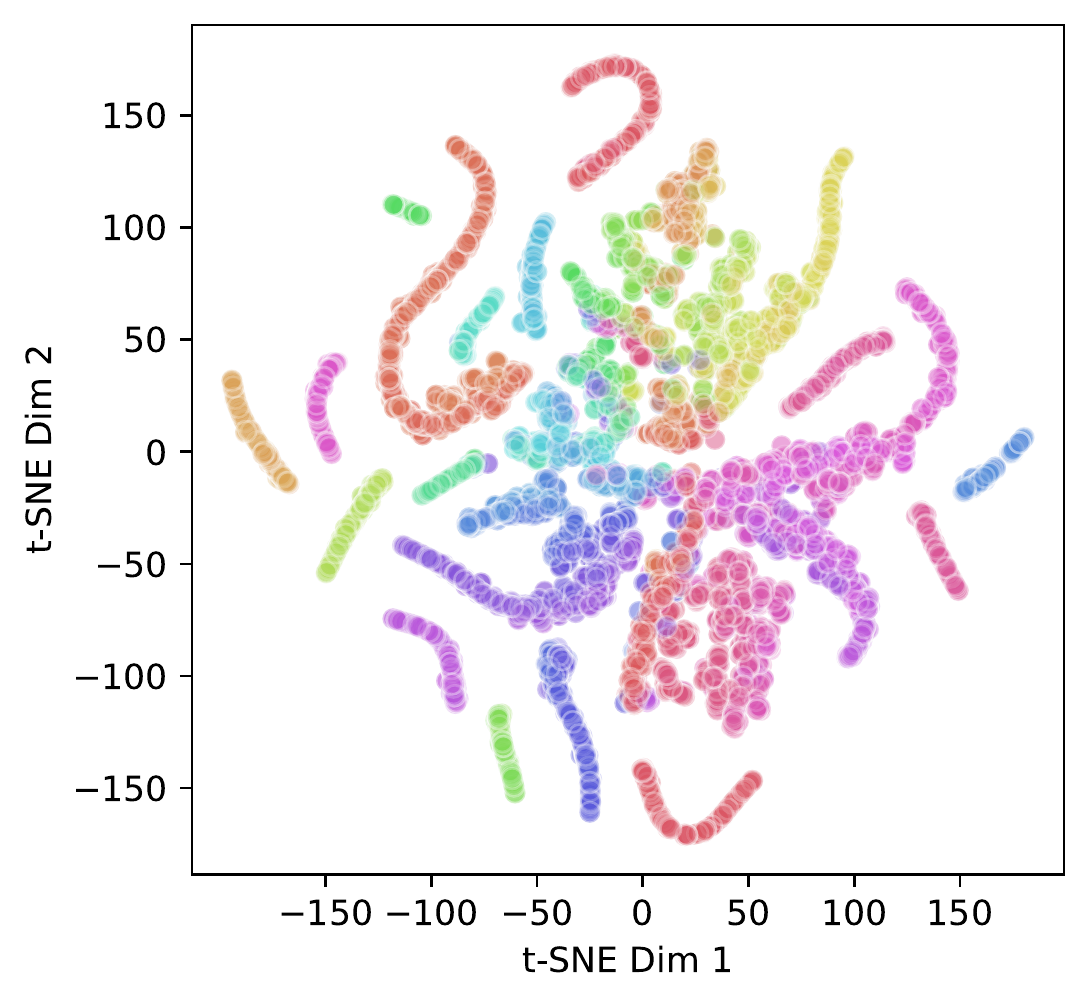}\label{figure:tsne_DFT_rosslyn}}
\hfill
\subfloat[DeepMIMO O1\_28, learned]{\includegraphics[width=0.33\columnwidth,trim=0.2cm 0.2cm 0.2cm 0.2cm,clip]{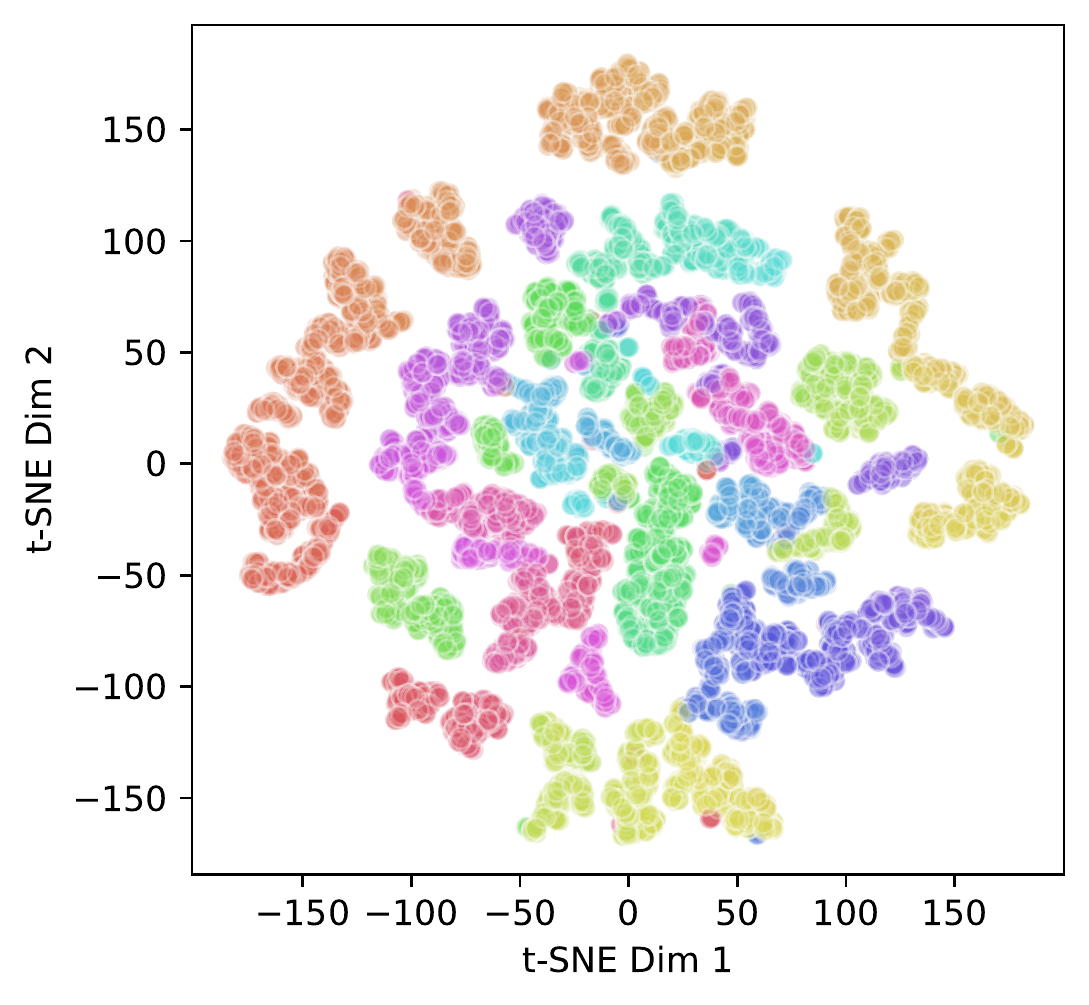}\label{figure:tsne_learned_O28}}
\subfloat[DeepMIMO O1\_28, AMCF]{\includegraphics[width=0.33\columnwidth,trim=0.2cm 0.2cm 0.2cm 0.2cm,clip]{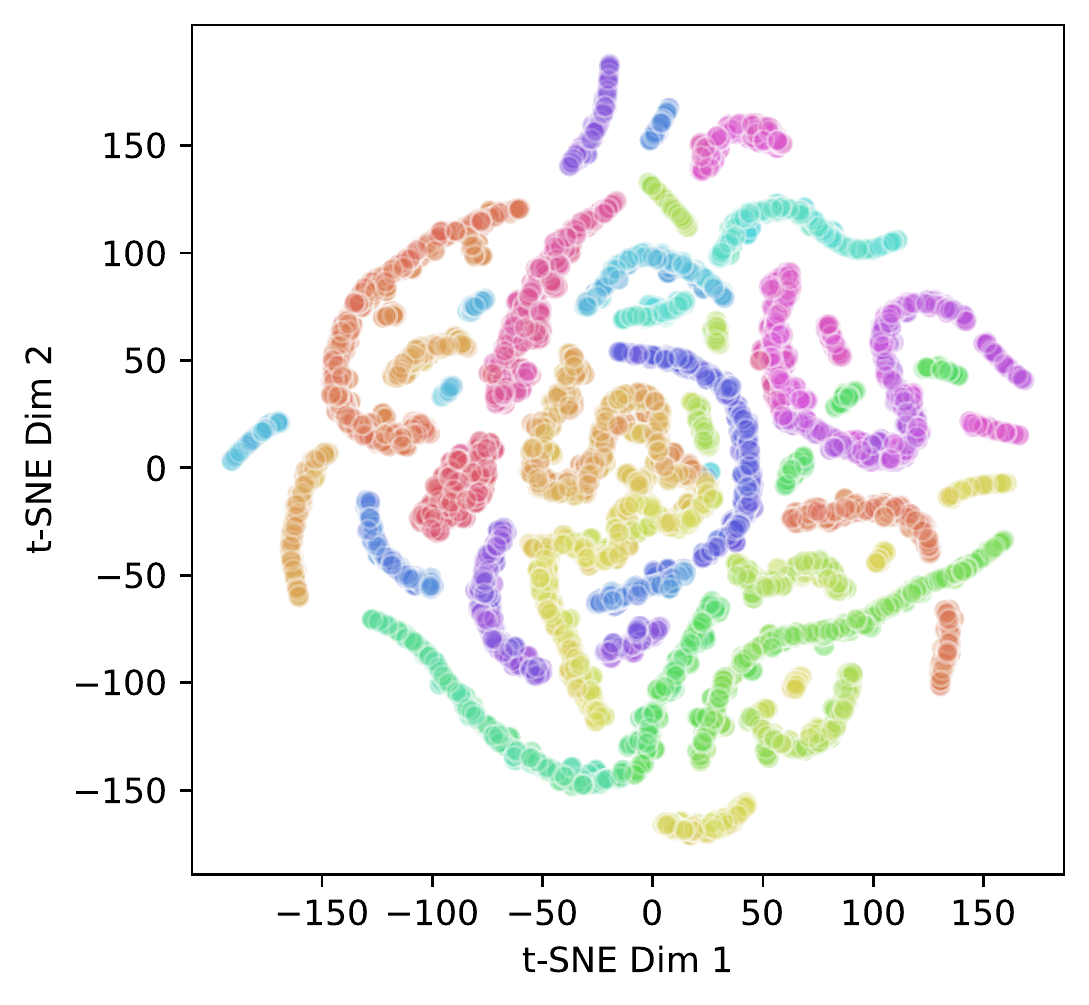}\label{figure:tsne_AMCF_O28}}
\subfloat[DeepMIMO O1\_28, DFT]{\includegraphics[width=0.33\columnwidth,trim=0.2cm 0.2cm 0.2cm 0.2cm,clip]{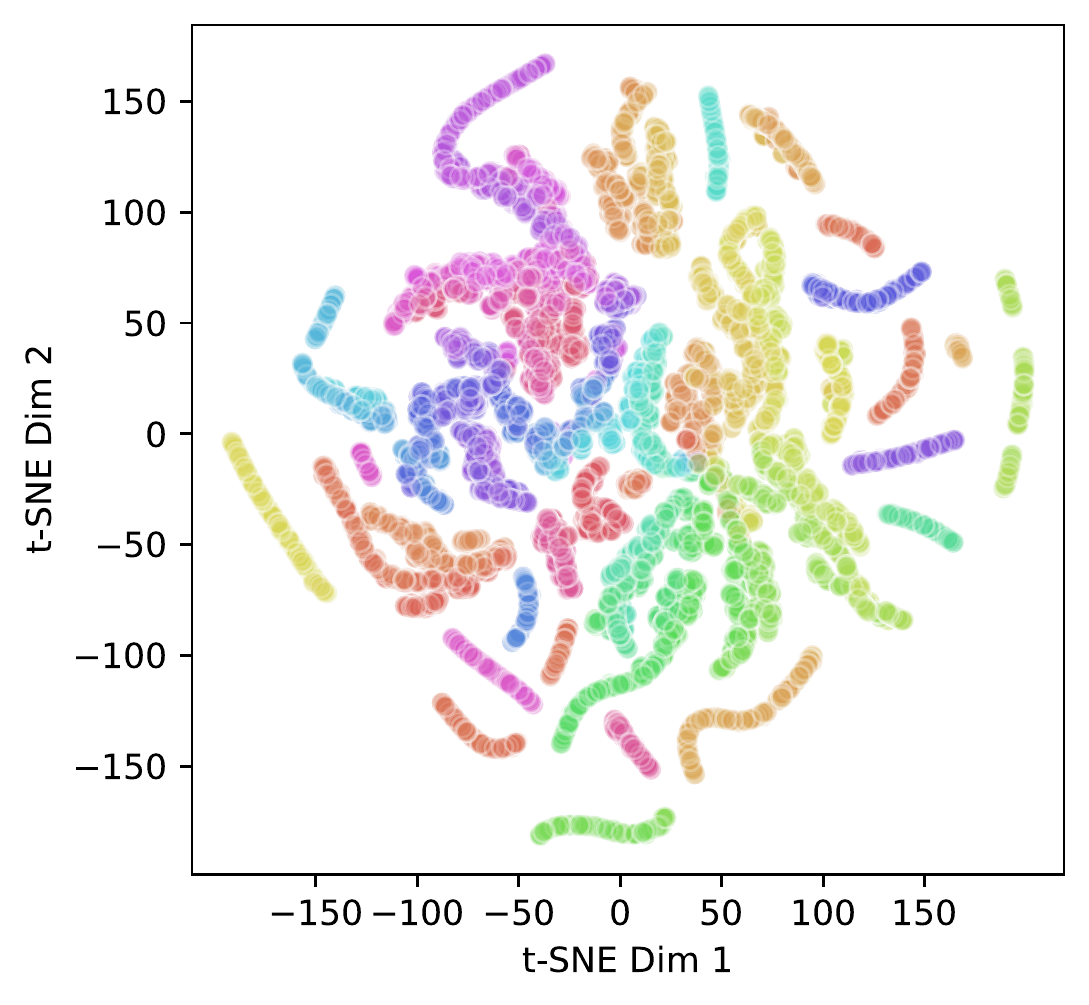}\label{figure:tsne_DFT_O28}}
\caption{t-SNE visualization of the received signal power of different probing codebooks with $N_{\mathbf{W}}=16$ beams.}\label{figure:tsne}
\end{figure*}

To further visualize the clustering effect of the probing codebooks, 2-D embeddings of the probing codebook measurements $\mathbf{x}$ are learned using the \ac{t-SNE} algorithm. The \ac{t-SNE} algorithm \cite{van2008tsne} is commonly used to learn low-dimensional embeddings of high-dimensional data while preserving its distribution in the high-dimensional space, so that similar data samples are more likely to be closer together and dissimilar ones are more likely to be farther apart in the embedding space. Two environments -- Rosslyn and DeepMIMO O1\_28 -- are selected as case studies, and their \ac{t-SNE} visualizations are shown in Fig. \ref{figure:tsne}. With the \ac{AMCF} and the \ac{DFT} probing codebooks, the clusters have elongated and twisted shapes. With the learned codebook, the data samples within each cluster are more tightly packed. The learned probing codebook allows the channel realizations to form better-shaped clusters which will likely make the data easier to classify.  

\begin{table*}
\small
\centering
  \caption{Silhouette Coefficients of Beam Sweeping Received Signal Power}\label{table:silhouette_coefficients}
    \begin{tabular}{| c | c | c | c | c | c | c | c | c | c |}
    \hline
    \multirow{2}{*}{\textbf{Environment}} & \multirow{2}{*}{\makecell{\textbf{Probing} \\\textbf{codebook}}} & \multicolumn{8}{c|}{\textbf{Probing codebook size}}\\ \cline{3-10}
     & & \textbf{6} & \textbf{8} & \textbf{10} & \textbf{12} & \textbf{14} & \textbf{16} & \textbf{18} & \textbf{20} \\ \hline
    \multirow{3}{*}{Rosslyn} 
    & learned & -0.179 & -0.144 & -0.118 & -0.099 & -0.078 & -0.071 & -0.051 & -0.035 \\ \cline{2-10}
    & AMCF & -0.333 & -0.266 & -0.262 & -0.242 & -0.215 & -0.196 & -0.172 & -0.160 \\ \cline{2-10}
    & DFT & -0.376 & -0.404 & -0.320 & -0.281 & -0.231 & -0.231 & -0.189 & -0.199 \\ \hline
    \multirow{3}{*}{\makecell{DeepMIMO \\ O1\_28}} 
    & learned & -0.256 & -0.231 & -0.196 & -0.173 & -0.175 & -0.151 & -0.161 & -0.119 \\ \cline{2-10}
    & AMCF & -0.367 & -0.337 & -0.316 & -0.290 & -0.281 & -0.283 & -0.250 & -0.238 \\ \cline{2-10}
    & DFT & -0.426 & -0.380 & -0.353 & -0.340 & -0.329 & -0.302 & -0.272 & -0.292 \\ \hline
    \multirow{3}{*}{\makecell{DeepMIMO \\ I3}}
    & learned & -0.302 & -0.255 & -0.223 & -0.218 & -0.206 & -0.191 & -0.190 & -0.177 \\ \cline{2-10}
    & AMCF & -0.505 & -0.477 & -0.440 & -0.406 & -0.391 & -0.390 & -0.366 & -0.349 \\ \cline{2-10}
    & DFT & -0.503 & -0.480 & -0.416 & -0.393 & -0.385 & -0.399 & -0.361 & -0.364 \\ \hline
    \multirow{3}{*}{\makecell{DeepMIMO \\ O1\_28B}} 
    & learned & -0.597 & -0.584 & -0.588 & -0.573 & -0.573 & -0.572 & -0.571 & -0.571 \\ \cline{2-10}
    & AMCF & -0.623 & -0.618 & -0.616 & -0.612 & -0.608 & -0.611 & -0.604 & -0.601 \\ \cline{2-10}
    & DFT & -0.746 & -0.742 & -0.696 & -0.683 & -0.668 & -0.662 & -0.641 & -0.634 \\ \hline
    \end{tabular}
\end{table*}

\section{Conclusion}\label{section:conclusion}
We propose a \ac{mmWave} beam alignment method that uses \ac{ML} to predict the optimal narrow beam using measurements of a learned probing codebook. 
We design a \ac{NN} architecture that optimizes the site-specific probing codebook so that it can capture particular characteristics of the propagation environment.
After an offline training phase, operators can implement the learned probing codebook using an \ac{RF} chain and use its beam sweeping results to select an optimal narrow beam or a few candidate beams to try, which is compatible with the beam alignment framework in 5G.
The proposed method can outperform hierarchical beam search baselines and even the exhaustive beam search, particularly in challenging environments with \ac{NLOS} \acp{UE}, while significantly reducing the beam sweeping overhead.
We also provide an explanation of why the learned probing codebook is beneficial to the beam alignment task through the lens of data clustering and representation learning. 
The proposed method uses channel information during its offline training phase. Future works may consider beam prediction without offline training or explicit channel knowledge. The complex-\ac{NN} architecture may also be extended to consider hybrid \ac{BF}. The extension to receive beam alignment on the \ac{UE} side is another promising direction.
 
\section{Acknowledgements}
The authors thank V. Va, A. Ali and B.L. Ng from Samsung Research America for their valuable feedback and discussion. 
\bibliographystyle{IEEEtran}
\bibliography{refs}

\end{document}

%% file: acronyms.tex
\DeclareAcronym{3GPP}{
  short=3GPP,
  long=3rd generation partnership project
}
\DeclareAcronym{ADC}{
  short=ADC,
  long=analog-to-digital converter
}
\DeclareAcronym{AMP}{
  short=AMP,
  long=approximate message passing
}
\DeclareAcronym{ANN}{
  short=ANN,
  long=artificial neural network
}
\DeclareAcronym{AoA}{
  short=AoA,
  long=angle-of-arrival
}
\DeclareAcronym{AoD}{
  short=AoD,
  long=angle-of-departure
}
\DeclareAcronym{APS}{
  short=APS,
  long=azimuth power spectrum
}
\DeclareAcronym{AR}{
  short=AR,
  long=augmented reality
}
\DeclareAcronym{AV}{
  short=AV,
  long=autonomous vehicle
}
\DeclareAcronym{BM}{
  short=BM,
  long=beam management
}
\DeclareAcronym{BS}{
  short=BS,
  long=base station
}
\DeclareAcronym{BSM}{
  short=BSM,
  long=basic safety message
}
\DeclareAcronym{CDF}{
  short=CDF,
  long=cumulative distribution function
}
\DeclareAcronym{CP}{
  short=CP,
  long=cyclic-prefix
}
\DeclareAcronym{CSI-RS}{
  short=CSI-RS,
  long=Channel State Information Reference Signal
}
\DeclareAcronym{DFT}{
  short=DFT,
  long=discrete Fourier transform
}
\DeclareAcronym{DL}{
  short=DL,
  long=downlink
}
\DeclareAcronym{EKF}{
  short=EKF,
  long=extended Kalman filter
}
\DeclareAcronym{DSRC}{
  short=DSRC,
  long=dedicated short-range communication
}
\DeclareAcronym{FDD}{
  short=FDD,
  long=frequency division duplex
}
\DeclareAcronym{FMCW}{
  short=FMCW,
  long=frequency modulated continuous wave
}
\DeclareAcronym{FoV}{
  short=FoV,
  long=field-of-view
}
\DeclareAcronym{GNSS}{
  short=GNSS,
  long=global navigation satellite system
}
\DeclareAcronym{IA}{
  short=IA,
  long=initial access
}
\DeclareAcronym{IMU}{
  short=IMU,
  long=inertial measurement unit
}
\DeclareAcronym{lidar}{
  short=lidar,
  long=light detection and ranging
}
\DeclareAcronym{LOS}{
  short=LOS,
  long=line-of-sight
}
\DeclareAcronym{LPF}{
  short=LPF,
  long=low pass filter
}
\DeclareAcronym{LTE}{
  short=LTE,
  long=long term evolution
}
\DeclareAcronym{MIMO}{
  short=MIMO,
  long=multiple-input multiple-output
}
\DeclareAcronym{ML}{
  short=ML,
  long=machine learning
}
\DeclareAcronym{mmWave}{
  short=mmWave,
  long=millimeter wave
}
\DeclareAcronym{MRR}{
  short=MRR,
  long=medium range radar
}
\DeclareAcronym{NLOS}{
  short=NLOS,
  long=non-line-of-sight
}
\DeclareAcronym{NB}{
  short=NB,
  long=narrow beam
}
\DeclareAcronym{NR}{
  short=NR,
  long=new radio
}
\DeclareAcronym{OFDM}{
  short=OFDM,
  long=orthogonal frequency-division multiplexing
}
\DeclareAcronym{ppm}{
  short=ppm,
  long=parts-per-million
}
\DeclareAcronym{PF}{
  short=PF,
  long=particle filter
}
\DeclareAcronym{RMS}{
  short=RMS,
  long=root-mean-square
}
\DeclareAcronym{RPE}{
  short=RPE,
  long=relative precoding efficiency
}
\DeclareAcronym{RS}{
  short=RS,
  long=reference signal
}
\DeclareAcronym{RSRP}{
  short=RSRP,
  long=reference signal received power
}
\DeclareAcronym{RSU}{
  short=RSU,
  long=roadside unit
}
\DeclareAcronym{SCS}{
  short=SCS,
  long=subcarrier spacing
}
\DeclareAcronym{SNR}{
  short=SNR,
  long=signal-to-noise ratio
}
\DeclareAcronym{SRS}{
  short=SRS,
  long=Sounding Reference Signal
}

\DeclareAcronym{SSB}{
  short=SSB,
  long=Synchronization Signal Block
}
\DeclareAcronym{THz}{
  short=THz,
  long=terahertz
}
\DeclareAcronym{UAV}{
  short=UAV,
  long=unmanned aerial vehicle
}
\DeclareAcronym{UE}{
  short=UE,
  long=user equipment
}
\DeclareAcronym{UKF}{
  short=UKF,
  long=unscented Kalman filter
}
\DeclareAcronym{UL}{
  short=UL,
  long=uplink
}
\DeclareAcronym{ULA}{
  short=ULA,
  long=uniform linear array
}
\DeclareAcronym{V2I}{
  short=V2I,
  long=vehicle-to-infrastructure
}
\DeclareAcronym{V2V}{
  short=V2V,
  long=vehicle-to-vehicle
}
\DeclareAcronym{V2X}{
  short=V2X,
  long=vehicle-to-everything
}
\DeclareAcronym{VR}{
  short=VR,
  long=virtual reality
}
\DeclareAcronym{VRU}{
  short=VRU,
  long=vulnerable road user
}
\DeclareAcronym{WB}{
  short=WB,
  long=wide beam
}
\DeclareAcronym{BF}{
  short=BF,
  long=beamforming
}
\DeclareAcronym{CS}{
  short=CS,
  long=compressive sensing
}
\DeclareAcronym{NN}{
  short=NN,
  long=neural network
}
\DeclareAcronym{RF}{
  short=RF,
  long=radio frequency
}
\DeclareAcronym{MISO}{
  short=MISO,
  long=multiple-input single-output
}
\DeclareAcronym{SIMO}{
  short=MISO,
  long=single-input multiple-output
}
\DeclareAcronym{MLP}{
  short=MLP,
  long=multilayer perceptron
}
\DeclareAcronym{AMCF}{
  short=AMCF,
  long=alternative minimization method with a closed-form expression
}
\DeclareAcronym{AWGN}{
  short=AWGN,
  long=additive white Gaussian noise
}
\DeclareAcronym{t-SNE}{
  short=t-SNE,
  long=t-distributed stochastic neighbor embedding
}
\DeclareAcronym{TDD}{
  short=TDD,
  long=time division duplex
}
\DeclareAcronym{mMTC}{
  short=mMTC,
  long=massive Machine Type Communications
}